\documentclass[11pt,a4paper]{article}
\pdfoutput=1
\usepackage{jheppub}
\usepackage{amsmath}
\usepackage{dsfont}
\usepackage{MnSymbol}
\usepackage{ulem}
\usepackage{array}
\usepackage{multirow}
\usepackage{hhline}
\usepackage{hyperref}

\makeatletter\renewcommand{\@biblabel}[1]{#1.}\makeatother

\def\be{\begin{equation}}
\def\ee{\end{equation}}

\preprint{
{\small{\textsf{}}}}

\title{Solving $q$-Virasoro constraints}

\author[a]{Rebecca Lodin,}
\author[a,b,c,d]{Aleksandr Popolitov,}
\author[d,e,f,1]{Shamil Shakirov%
\note{On leave from Harvard University}}
\author[a]{and Maxim Zabzine}

\affiliation[a]{Department of Physics and Astronomy, Uppsala University,\\
Box 516, SE-75120 Uppsala, Sweden.}

\affiliation[b]{Moscow Institute for Physics and Technology, Dolgoprudny, Russia}

\affiliation[c]{ITEP, Moscow 117218, Russia}

\affiliation[d]{Institute for Information Transmission Problems, Moscow 127994, Russia}

\affiliation[e]{Society of Fellows, Harvard University, Cambridge, MA 02138, USA}

\affiliation[f]{Mathematical Sciences Research Institute, Berkeley, CA 94720, USA}

\emailAdd{rebecca.lodin@physics.uu.se}
\emailAdd{aleksandr.popolitov@physics.uu.se}
\emailAdd{shakirov@fas.harvard.edu}
\emailAdd{maxim.zabzine@physics.uu.se}

\abstract{
We show how $q$-Virasoro constraints can be derived for a large class of $(q,t)$-deformed eigenvalue matrix models by an elementary trick of inserting certain $q$-difference operators under the integral, in complete analogy with full-derivative insertions for $\beta$-ensembles. From free field point of view the models considered have zero momentum of the highest weight, which leads to an extra constraint $T_{-1} \mathcal{Z} = 0$. We then show how to solve these $q$-Virasoro constraints recursively and comment on the possible applications for gauge theories, for instance calculation of (supersymmetric) Wilson loop averages in gauge theories on 
$D^2 \times S^1$ and $S^3$.}

\DeclareMathOperator*\Res{Res}

\begin{document}
\today

\maketitle

\flushbottom

\section{Introduction}

Over the last 40 years matrix models have attracted enormous attention in mathematical physics. Besides their direct applications, matrix models provide a useful playground for quantum field theory and as such they have been extensively studied (see \cite{Morozov:1994hh} and references therein).

The simplest examples of matrix models are ordinary integrals over the space of finite dimensional matrices. However, the {\it real} interest is in more complicated, deformed, examples, where the link to the integrals over matrices is less direct or even completely lost. The central question is then: which (hidden) structures do survive the deformation \cite{Morozov:2005mz}? The recently renewed interest in deformed matrix models is fueled by applications in gauge theories (calculations using localization technique \cite{Pestun:2016zxk} often lead to an effective description in terms of matrix models), as well as the theory of symmetric polynomials and quantum groups \cite{Macdonald,frenkel1996}. The deformation that is immediately relevant to supersymmetric gauge theories is the so-called $(q,t)$-deformation. In this paper we show that certain standard tools available in the non-deformed cases can also, in fact, be used in the $(q,t)$-deformed case.

The central object in any matrix model is the partition function $\mathcal{Z}$, which is, usually but not always,
a formal power series in infinitely many "time" variables $\{t_k\}$.
Coefficients of $\mathcal{Z}$ with respect to $\{t_k\}$ are called correlators and one of the questions for each given class of models is how to find these correlators, ideally in a fast and efficient way.

In the case of non-deformed models, exemplified by the Hermitian 1-matrix model \cite[Section~3.2]{Morozov:1994hh}, there are many answers to the question of how to find correlators. One way is to use the character decomposition formulas, which are also conjecturally available in the deformed case \cite{Morozov:2018eiq}. Another method is to derive Ward identities and use them as equations for correlators. Yet another way is to derive the so-called loop equations from the Ward identities and further recast them into the algebrogeometric form known as the spectral curve topological recursion \cite{Eynard:2007kz,Alexandrov:2006qx,Alexandrov:2009gn,Orantin:2008hq}. All these different ways, which can be thought of as manifestations of different hidden structures of the model, are straightforwardly linked
to one another in the non-deformed case
(and in the case of Hermitian Gaussian 1-matrix model it is especially easy to trace all the connections).
However, after $(q,t)$-deformation, each of the hidden structures changes in its own peculiar way.
As a result, relations become less apparent and so each structure needs to be studied by itself.

Currently, some of the deformed structures are understood more and some are understood less.
Character expansion is confirmed by computer experiments, yet its derivation from first principles is a challenging open problem
\cite{Morozov:2018eiq}.
The $(q,t)$-generalization of the topological recursion is completely unknown
(but, on the other hand, topological recursion for $\beta$-ensembles is well-understood, see \cite{ChekhovEynardMarshal2010}). It is not even clear how to distinguish contributions of different genera in a correlator.
The situation with the $(q,t)$-deformed Ward identities, which are the main focus of this paper, is the following.
In the non-deformed case, there are two different ways to derive the Ward identities.
The first one, which is more conceptual as it exhibits connection between matrix models and conformal field theories, relies on representing the partition function of a given model as an average in some free field theory \cite{Nedelin:2016gwu}.
The second one, which looks more like a clever trick, consists of insertion of a full derivative under the integral \cite[Section~2.1]{Morozov:1994hh}.
The $(q,t)$-version of the first derivation is well known \cite{Nedelin:2016gwu}.
From it, we know that for a large class of $(q,t)$-deformed models Ward identities form the so-called $q$-Virasoro algebra
\cite{Shiraishi:1995rp}.
However, the derivation of the $q$-Virasoro constraints using the insertion of a full derivative (which should be substituted with a full $q$-difference operator) is not known in general, but there are some exceptions. In \cite{Zenkevich:2014lca} it is done in the case of $(q,t)$-deformed Selberg integral.
The $(q,t)$-deformed Selberg integral is a very special point in the space of all $(q,t)$-deformed matrix models as it is the {\it only} matrix model for which the character expansion is proved \cite[Chapter~VI, Section~9, Example~3]{Macdonald}.
The sketch of the generic derivation can be found in \cite[Section~2.1]{Mironov:2016yue}, though the details are not spelled out.
In this paper we do derive the $q$-Virasoro constraints using the insertion of a full difference operator (details are in Section~\ref{differenceoperator}) for a large class of models (defined in Section~\ref{definition}).

As it often happens, working out the details brings interesting surprises.
In order for the equations to be sufficient to determine all correlators in the simplest case of Gaussian $(q,t)$-deformed model, we need to consider one ``additional" equation
(see Section~\ref{sec:special-additional-case}),
which is the $(q,t)$-analogue of the string equation \cite[Introduction]{Alexandrov:2009gn}.
When we spell out what this equation means in the language of free fields, it turns out to be nothing but the condition $T_{-1} \mathcal{Z} = 0$ (usually the partition function is annihilated only by positive $q$-Virasoro generators, see Section~\ref{matrixmodel}).
This is an extra equation that is valid only for partition functions with zero vacuum momentum.

Having derived $q$-Virasoro constraints we proceed to solving them.
Namely, we describe, how to use them to find all correlators of the model.
In the non-deformed case, thanks to the simple form of the Virasoro constraints (they are second order differential operators)
it is obvious how to convert them into efficient recursive procedure, that allows to calculate each given correlator in polynomial time. The form of $q$-Virasoro constraints is more complicated and it is far less obvious
how to use them to get an efficient recursion.
Still, we find such recursion in Section~\ref{recursive} (see equation \eqref{eq:q-vir-correlators-explicit}).

This efficient procedure for evaluation of correlators is of immediate interest for gauge theory applications. In gauge theories one is interested in supersymmetric Wilson loop averages. On the $(q,t)$-matrix model side, these averages correspond to the correlators of Schur functions (even though the Macdonald functions are more natural from the $(q,t)$ point of view). Each Schur function is a finite linear combination of power sum monomials, therefore, one can use the recursion procedure of Section~\ref{recursive},
to evaluate any supersymmetric Wilson loop in any gauge theory,
that corresponds to some matrix model from the class we define in Section~\ref{definition}.

Last but not least, from the free field derivation one can see that $(q,t)$-deformed models have a vast hidden freedom which is not present in the non-deformed case: one can insert arbitrary $q$-constants under the integral. In our derivation using difference operators we also see that Ward identities for discrete and continuous Gaussian matrix models
(equations \eqref{eq:discrete-gaussian-mm} and \eqref{blockIntegral}, at the special choice of masses and Fayet-Iliopoulos parameter, respectively) are the same
(and hence the correlators are the same).
From the gauge theory point of view this looks like a non-trivial duality between the corresponding gauge theories.
So, the $q$-Virasoro viewpoint is potentially very fruitful in the search for dualities.

The rest of the paper is organized as follows. In Section~\ref{definition} we define the class of models we study and provide the two models we use as examples.
In Section~\ref{matrixmodel} we give the matrix model background and the free field realization of the $q$-Virasoro algebra,
and in Section~\ref{differenceoperator} we move on to the derivation of the constraint equations which the models have to satisfy.
In Section~\ref{recursive} we consider the recursive solution to correlators and in Section~\ref{gaugetheory} we see how this can be applied to gauge theory.
Finally in Section~\ref{conclusion} we conclude and suggest further interesting directions of investigation.
Definitions of special functions and details of computations are left to the appendices.

\section{Definition of the class of models} \label{definition}
Let us consider a class of models of dimension $N$ where the partition function is given by
\begin{equation} \label{generatingfn}
\mathcal{Z}\{t_k\} = \oint_{C_i} \prod_{i=1}^N \frac{dx_i}{x_i} \ F(\underline{x})~,
\end{equation}
for some appropriately chosen contours $\{C_i\}$, with integrand
\begin{equation} \label{Fgeneratingfn}
F(\underline{x}) = \prod_{k \neq l} \frac{(x_k/x_l;q)_\infty }{(tx_k/x_l;q)_\infty} \ e^{\sum_{k=1}^\infty t_k \sum_{i=1}^N x_i^k} \ \prod_{i=1}^N c_q(x_i)~.
\end{equation}
Here $(x;q)_\infty = \prod_{k=0}^{\infty}(1-x q^k)$ is the $q$-Pochhammer symbol and $c_q$ is of the form
\begin{equation} \label{cqgeneral}
  c_q(x) \ = \ x^{\sqrt{\beta}(\alpha+\sqrt{\beta}N-Q_\beta)} \lambda_q(x) \
  \prod_{f=1}^{N_f} \frac{(q x \overline{m}_f;q)_\infty }{(x m_f;q)_\infty }~,
\end{equation}
where $\alpha \in \mathbb{C}$ (later to be interpreted as momentum of CFT states, see \eqref{state}) and $\lambda_q(x)$ is a $q$-constant\footnote{A $q$-constant is a function $g(\underline{x};q)$ with the property $g(x_1,\dots , qx_i, \dots , x_N;q)=g(\underline{x};q)$.}. The parameters $t,q,\beta \in \mathbb{C}$ are related via $t=q^\beta$, while $Q_\beta=\sqrt{\beta}-1/\sqrt{\beta}$.

This is a natural $(q,t)$-generalization of eigenvalue matrix models and of $\beta$-ensembles: \linebreak indeed, the first factor in (\ref{Fgeneratingfn}) plays the role of $(q,t)$-deformed Vandermonde measure, the second introduces time-variables $\{t_k\}$ and the third encodes the common potential for the "eigenvalues" $x_i$. In gauge theory applications $N_f$ is the number of (anti-)fundamental masses and the parameters of the potential $m_f$ are the respective fugacities. Note that the product
$\prod_{f=1}^{N_f} \frac{(q x \overline{m}_f;q)_\infty }{(x m_f;q)_\infty }$ can be generated by suitable shifts in $\{t_k\}$.

Of course, this is not the first time this class of models makes an appearance in the literature. For example, in pure mathematics they appeared in the context of constant term evaluations and Macdonald conjectures (see \cite{Warnaar2007} for a review). In the physical context they played an important role in the description of AGT dualities \cite{Mironov:2011dk}, trialities between conformal field theories and gauge theories in three and five dimensions \cite{Aganagic:2013tta, Aganagic:2014oia} and more recently
hidden symmetries of network-type matrix models and their relation to DIM algebras \cite{Mironov:2016yue}.

\begin{figure}[h]
\includegraphics[scale=0.9]{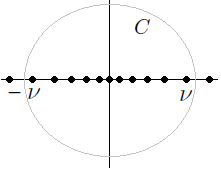}
\centering
\caption{Illustration of integration contour} \label{contour_nu}
\end{figure}

Within this class of models there are two examples that we will frequently use for illustration purposes.
The first one is the discrete $(q,t)$-Gaussian model, obtained by taking $N_f=2$ with two opposite masses $\pm m$ having the special values $\pm m=\pm q(1-q)^{\frac{1}{2}}$. In the limit  $q\rightarrow 1$ the product of the two $q$-Pochhammer symbols becomes $e^{-\frac{x^2}{2}}$ and the standard Gaussian potential is recovered. This justifies the name $(q,t)$-Gaussian model as the simplest deformation of the Gaussian\footnote{For $N_f>2$ this appears to give a natural $(q,t)$-generalization of non-Gaussian matrix models, whose characteristic non-Gaussian features (such as contour dependence and multi-cut phases) would be very interesting to investigate in the future.} model. This model was axiomatically defined in \cite{Morozov:2018eiq} via its Macdonald averages,
but here we give the explicit integral representation.
The function $c_q(x_i)$ for the $(q,t)$-Gaussian model can be written as
\begin{equation} \label{cqGaussian}
c_q(x) = x^{\sqrt{\beta}(\sqrt{\beta}N-Q_\beta)}(x^2q^2\nu^{-2};q^2)_\infty (1-q)\Bigg\{ -\frac{(q;q)_\infty ^2}{\theta (q^\lambda;q)} \left[ f_\lambda(x/\nu;q)-f_\lambda(-x/\nu;q) \right] \Bigg \}
\end{equation}
where $\theta(x;q) = (x;q)_\infty (qx^{-1};q)_\infty$ is the Jacobi theta function, $\nu =(1-q)^{-\frac{1}{2}}$, and
\begin{equation}
f_\lambda(x;q) = x^\lambda \ \frac{\theta(q^\lambda x;q)}{\theta(x;q)}~,
\end{equation}
is a particular $q$-constant. Alternatively, the same model can be expressed in a more concise form using the notion of a Jackson $q$-integral -- a discrete $q$-analogue \eqref{qintegral} of an integral -- via summing over residues \eqref{qintegralpoles}, provided that contours are chosen as in figure \ref{contour_nu}:
\begin{equation} \label{eq:discrete-gaussian-mm}
\mathcal{Z}_{(q,t)\textrm{-Gaussian}}\{t_k\} = \int_{-\nu}^\nu \prod_{i=1}^N d_qx _i \prod_{j=1}^N (x_j^2q^2\nu^{-2};q^2)_\infty \prod_{k \neq l =1}^N x_l^\beta \frac{(x_k/x_l;q)_\infty }{(tx_k/x_l;q)_\infty} e^{\sum_{k=1}^\infty t_k \sum_{i=1}^N x_i^k}~.
\end{equation}
These two descriptions -- continuous and discrete -- provide two useful and complementary perspectives on one and the same model.

The second example that we consider is the 3d $\mathcal{N} =2$ gauge theory partition function on $D^2 \times S^1$ where the non-trivial decomposition was found in \cite{Beem:2012mb} and the $q$-Virasoro interpretation was found in \cite{Aganagic:2013tta} and extended in \cite{Nedelin:2016gwu}. The holomorphic block integral (omitting factors of $2\pi i $) for this gauge theory takes the form
\begin{equation} \label{blockIntegral}
\mathcal{B}_c^{3d} = \oint_c \prod_{i=1}^N \frac{dx_i}{x_i}  \prod_{k \neq l =1}^N \frac{(x_k/x_l;q)_\infty }{(m_ax_k/x_l;q)_\infty } \prod_{j=1}^Nx_j^{\kappa_1} \prod_{f=1}^{N_f} \frac{(qx_j\overline{m}_f;q)_\infty }{(x_jm_f;q)_\infty }~,
\end{equation}
where the integration is over middle dimensional cycles $c$ in $(\mathbb{C}^\times)^{\textrm{rkG}}$, $m_a$ is the adjoint fugacity, $m_f$ and $\overline{m}_f$ are the fugacities for the fundamental and anti-fundamental and $\kappa_1$ is the Fayet-Iliopoulos parameter. As given in \citep{Nedelin:2016gwu} we have the identification $\kappa_1 = \sqrt{\beta}(\alpha_0+\sqrt{\beta}N-Q_\beta)$ (for some initial momenta $\alpha_0 \in \mathbb{C}$) between the gauge theory and the $q$-Virasoro side, and the model is then reproduced by
\begin{equation} \label{cqGauge}
\lambda_q(x) = 1~,
\end{equation}
with the generating function
\begin{equation}
\mathcal{Z}_{D^2 \times S^1}\{t_k\} = \oint_c \prod_{i=1}^N \frac{dx_i}{x_i} \prod_{k \neq l=1}^N \frac{(x_k/x_l;q)_\infty }{(tx_k/x_l;q)_\infty} \ \prod_{j=1}^N \prod_{f=1}^{N_f} \frac{(qx_j\overline{m}_f;q)_\infty }{(x_jm_f;q)_\infty } \ e^{\sum_{k=1}^\infty t_k \sum_{i=1}^N x_i^k+\kappa_1\sum_{i=1}^N\ln x_i}~.
\end{equation}
These two examples are used in the paper to illustrate various points, though of course everything holds
for the class of models defined by equations \eqref{generatingfn}, \eqref{Fgeneratingfn}, \eqref{cqgeneral}, with certain reservations. Here we summarize these restrictions on the parameters of the model, since they are otherwise scattered
throughout Section~\ref{differenceoperator}, with references to the places they occur:
\begin{itemize}
\item The function $c_q(x)$ doesn't have poles between eigenvalue integration contours $C_i$ and their rescaled versions $C_i / q$
  (see Section~\ref{sec:the-generic-case}, after equation \eqref{nopoles});
\item The ratio $c_q(1/x)/c_q(q/x)$ should have no poles except maybe at $0$ and $\infty$
  (see equation \eqref{eq:contour-transformation} and explanations around it);
\item Parameters of the model satisfy $|m_f| < |q|$, $|q| < 1$, $|t| < |q|$
  (Section~\ref{sec:the-generic-case}, after equation \eqref{nopoles});
\item Moreover, in order to match with the simple representation of $q$-Virasoro,
  dependence of partition function $\mathcal{Z}$ on $m_f$ has to be studied formally
  (see explanation after equation \eqref{eq:contour-transformation}),
\item Momentum $\alpha$ should be zero (see equation \eqref{eq:no-potential-virasoro} and explanations around it).
\end{itemize}

We will now move on to review the properties of the free field realization of $q$-Virasoro algebra that will be needed to understand the above constraints.

\section{$(q,t)$-deformed Virasoro matrix model and free fields} \label{matrixmodel}
In this section we recall the necessary facts about the $(q,t)$-deformed eigenvalue models and their free field realization,
where details can be found in \cite{Morozov:1994hh,Nedelin:2016gwu,Awata:2010yy,Odake:1999un}. It can be noted that any model of the form \eqref{generatingfn} can be generated using the free field representation given below, and so there is no discrepancy between the different point of views.
\\
\\
The $q$-Virasoro algebra is given by the set of generators $\{T_{n}, n \in \mathbb{Z}\}$,
subject to relations
\begin{equation}
[T_n, T_m] = -\sum_{l>0} f_l \left( T_{n-l}T_{m+l}-T_{m-l}T_{n+l} \right) -\frac{(1-q)(1-t^{-1})}{(1-p)}(p^n-p^{-n})\delta_{n+m,0}
\end{equation}
or equivalently, in terms of generating currents
\begin{equation}
f(w/z) T(z) T(w) - f(z/w) T(w) T(z) = -\dfrac{(1-q)(1-t^{-1})}{1-p} \ \big\{ \ \delta(pw/z) - \delta(p^{-1}w/z) \ \big\}
\end{equation}
where
\begin{equation}
T(z)=\sum_{n \in \mathbb{Z}}T_n z^{-n}, \ \ \ f(z) = \sum\limits_{l \geq 0} f_l \ z^l = e^{\sum\limits_{n \geq 0} \frac{(1-q^n)(1-t^{-n})}{n(1+p^n)}z^n}, \ \ \ \delta(z) = \sum_{n \in \mathbb{Z}} z^{n}
\end{equation}
and the parameters $p, q, t \in \mathbb{C}$ and $p=q/t$. The free field representation of the $q$-Virasoro algebra is given in terms of the following Heisenberg oscillators
\begin{equation}
\begin{split}
[\mathsf{a}_n, \mathsf{a}_m] = & \frac{1}{n}(q^{\frac{n}{2}}-q^{-\frac{n}{2}})(t^{\frac{n}{2}}-t^{-\frac{n}{2}})(p^{\frac{n}{2}}+p^{-\frac{n}{2}})\delta_{n+m,0}~,~~~~n,m \in \mathbb{Z} \backslash \{0\}~,\\
[\mathsf{P}, \mathsf{Q}] =&  2~,
\end{split}
\end{equation}
using which the stress tensor current takes form
\begin{equation}
T(z) = \sum_{\sigma = \pm 1}\Lambda_\sigma(z) = \sum_{\sigma = \pm 1} :e^{\sigma \sum_{n \neq 0}\frac{z^{-n}}{(1+p^{-\sigma n})}\mathsf{a}_n}:q^{\sigma \frac{\sqrt{\beta}}{2}\mathsf{P}}p^{\frac{\sigma}{2}}~.
\end{equation}
Explicit expressions for the generators are given by
\begin{equation}\label{generators}
T_n =\begin{cases}
& \sum_{\sigma = \pm 1}q^{\sigma \frac{\sqrt{\beta}}{2}\mathsf{P}}p^{\frac{\sigma }{2}}\sum_{k\geq 0} s_{k}(\{A_{-k}^{(\sigma)}\})s_{n+k}(\{A_{n+k}^{(\sigma)}\})~,~~~~n\geq 0\\
& \sum_{\sigma = \pm 1}q^{\sigma \frac{\sqrt{\beta}}{2}\mathsf{P}}p^{\frac{\sigma }{2}}\sum_{k\geq 0} s_{k-n}(\{A_{n-k}^{(\sigma)}\})s_{k}(\{A_{k}^{(\sigma)}\})~,~~~~n<0.
  \end{cases}
\end{equation}
Here we let
\begin{equation}
A_{n}^{(\sigma)}=\sigma \frac{\mathsf{a}_n|n|}{(1+p^{-\sigma n})}
\end{equation}
and
\begin{equation}
s_n(\{A_n\})= s_n(A_1,...,A_n)
\end{equation}
is the Schur polynomial in symmetric representation $[n]$, defined by \eqref{eq:symm-schur-poly} and in particular $s_0=1$ and $s_1(A_1)=A_1$. We will also use the time representation of the Heisenberg algebra in terms of time variables $\{t_k\}$:
\begin{equation} \label{eq:heisenberg-representation}
\begin{split}
\mathsf{a}_{-n} \simeq & (q^{\frac{n}{2}}-q^{-\frac{n}{2}})t_n~,~~~~ \mathsf{a}_{n} \simeq (t^{\frac{n}{2}}-t^{-\frac{n}{2}})(p^{\frac{n}{2}}+p^{-\frac{n}{2}})\frac{\partial}{\partial t_n}~,~~~~n\in \mathbb{Z}_{>0} \\
       { \mathsf{Q} \simeq  \sqrt{\beta} t_0}&~,
       ~~~~{\mathsf{P} \simeq 2\frac{1}{\sqrt{\beta}}\frac{\partial}{\partial t_0}~,~~~|\alpha \rangle=e^{\frac{\alpha}{2}\mathsf{Q}}|0\rangle \simeq e^{\sqrt{\beta} t_0 \frac{\alpha}{2}} \cdot 1}~.
\end{split}
\end{equation}
We can then define the screening current
\begin{equation}
\mathsf{S}(x) = :e^{-\sum_{n \neq 0}\frac{x^{-n}}{(q^{n/2}-q^{-n/2})}\mathsf{a}_n}:e^{\sqrt{\beta}\mathsf{Q}}x^{\sqrt{\beta}\mathsf{P}}~,
\end{equation}
which has the nice property of commuting with the generators up to a total $q$-derivative
\begin{equation}
[T_n, \mathsf{S}(x)]=\frac{\mathsf{O}_n(qx)-\mathsf{O}_n(x)}{x}
\end{equation}
for some operator $\mathsf{O}_n(x)$. Using this screening current we can then consider the state
\begin{equation} \label{state}
\mathcal{Z}\{t_k\} = \oint \prod_{i=1}^N \frac{dx_i}{x_i} \prod_{j=1}^N \mathsf{S}(x_j) |\alpha \rangle
\end{equation}
for some momentum $\alpha \in \mathbb{C}$ where the integration contours are chosen such that upon the insertion of a total derivative $\mathcal{D}$, the integral vanishes:
\begin{equation}
\oint \mathcal{D} \prod_{i=1}^N \frac{dx_i}{x_i} \prod_{j=1}^N \mathsf{S}(x_j) |\alpha \rangle =0~.
\end{equation}
Here it can be mentioned that this requirement can be shown to hold for a wider class of models which can have measures other than $\prod_{i=1}^N \frac{dx_i}{x_i}$, e.g.
 one can use either the Jackson integral or integral over $\prod_{i=1}^N \frac{dx_i}{x_i}  \lambda_q(x_i)$ with some $q$-constant $\lambda_q(x_i)$.
Since the vacuum $|\alpha\rangle$ is annihilated by positive $q$-Virasoro generators,
the state (partition function) $\mathcal{Z}\{t_k\}$ in equation \eqref{state} also satisfies the $q$-Virasoro constraints
\begin{equation} \label{qVirasoroConstraint}
T(z) \mathcal{Z}\{t_k\} = f(z)~~~~ \Rightarrow ~~~~ T_{n>0} \mathcal{Z}\{t_k\} = 0
\end{equation}
for some function $f(z)$ holomorphic at $z \rightarrow 0$.
Since the vacuum is an eigenvector of $T_0$ \cite{Shiraishi:1995rp}
\begin{equation}
T_0 | \alpha \rangle =\lambda_\alpha | \alpha \rangle~, \ \ \ \ \ \lambda_\alpha = p^{\frac{1}{2}} q^{\frac{\sqrt{\beta} \alpha}{2}} + p^{-\frac{1}{2}} q^{-\frac{\sqrt{\beta}\alpha}{2}}
\end{equation}
the partition function is also an eigenvector of $T_0$ with the same eigenvalue,
\begin{equation}
T_0 \mathcal{Z}\{t_k\} = \lambda_\alpha \mathcal{Z}\{t_k\}
\end{equation}
Additionally, restricting to the subset of states which has zero momentum, $\alpha =0$, we have an additional equation
(verified in Appendix~\ref{FreeField})
\begin{align}
T_{-1}\mathcal{Z}\{t_k\} =0~.
\end{align}

Two points deserve to be mentioned here. First, from the point of view of full $q$-derivative insertion,
the additional equation $T_{-1}\mathcal{Z}\{t_k\} =0$ arises quite naturally (see Section~\ref{sec:special-additional-case}).
The requirement of zero momentum there takes the form of the vanishing of the coefficient in front of some complicated term,
which is otherwise unmanageable. Once this term is gone, at least for $(q,t)$-Gaussian matrix model,
we can recursively determine all the correlators.

Second, from the point of view of the $q$-Virasoro representation theory, the condition $T_{-1}\mathcal{Z} =0$
(which can be rewritten as $[T_1, T_{-1}]\mathcal{Z} =0$) is necessary to have the so-called singular vector
at level 1. In a sense, it is the simplest irreducible representation of $q$-Virasoro.
It would be interesting to know, which matrix models (in a sense of explicit integral representation)
correspond to more complicated $q$-Virasoro representations.

\section{Ward identities via $q$-difference operator insertion} \label{differenceoperator}


We now proceed to the derivation of the Ward identities.
First, in Section~\ref{sec:q-diff-op} we define the $q$-difference operator that is to be inserted
under the integral. Then, in Section~\ref{sec:derivation-of-constraints} we gradually rewrite
the $q$-difference insertion as the action of shift operator in times $\{t_k\}$, which culminates in equation \eqref{eq18}
-- the main result of this section. Finally, we compare the resulting equation~\eqref{eq18} with the $q$-Virasoro equations
\eqref{qVirasoroConstraint} confirming that they match.
On the way we discover that in order to proceed we need to impose certain conditions on the parameters of the model, in particular
on the function $c_q(x)$ and on the integration contours for the eigenvalues.

\subsection{The $q$-difference operator} \label{sec:q-diff-op}


We want Ward identities to be a corollary of the fact that insertion of the suitably chosen
full $q$-difference operator under the integral vanishes, i.e. that
\begin{equation}\label{general_model}
\begin{split}
\oint_{C_1} \dots \oint_{C_i} \dots \oint_{C_N} \prod_{j=1}^N \frac{d x_j}{x_j}\sum_{i=1}^N \mathcal{D}_q (x_i)G_i(\underline{x}) F(\underline{x}) =  0~,
\end{split}
\end{equation}
where the $q$-difference operator is
\begin{equation}
\mathcal{D}_q(x_i)=(x_i^\dagger -1)
\end{equation}
with
\begin{equation} \label{dagger}
x_i^\dagger f(\underline{x}) = f(x_1, \dots , x_i/q,\dots , x_N)
\end{equation}
together with
\begin{equation}\label{firstG}
G_i(\underline{x}) =  \frac{1}{1-zx_i} \prod_{j \neq i} \frac{x_j-tx_i}{x_j-x_i}~,
\end{equation}
and $F(\underline{x})$ as given in \eqref{Fgeneratingfn}.
Here the dependence of $G_i(\underline{x})$ on $z$ is understood formally -- we will obtain an equation for each coefficient
in front of particular positive power of $z$.

\begin{figure}[h]
\includegraphics[scale=0.5]{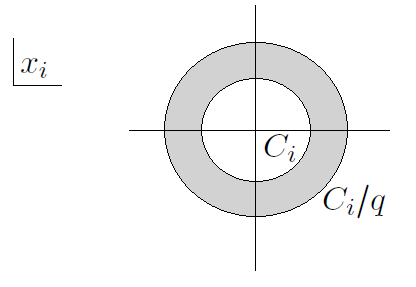}
\centering
\caption{Illustration of integration contour} \label{contour}
\end{figure}

For the equation \eqref{general_model} to hold, we must impose certain restrictions
on the integration contours and the potential term $c_q(x_i)$. Namely, let's consider terms in \eqref{general_model}
that contain operator $x^\dagger$. This operator can be traded for the change of the integration contour
as follows
\begin{equation} \label{eq:contour-shrink}
\begin{split}
\oint_{C_1} & \dots \oint_{C_i} \dots \oint_{C_N} \prod_{j=1}^N \frac{d x_j}{x_j}\sum_{i=1}^N x_i^\dagger G_i(\underline{x}) F(\underline{x}) \\
= & \sum_{i=1}^N \oint_{C_1} \dots \oint_{C_i} \dots \oint_{C_N} \prod_{j=1}^N \frac{d x_j}{x_j} G_i(x_1, \dots, x_i/q, \dots, x_N) F(x_1, \dots, x_i/q, \dots, x_N) \\
= & \sum_{i=1}^N \oint_{C_1} \dots \oint_{C_i/q}\dots \oint_{C_N} \prod_{j=1}^N \frac{d x_j}{x_j} G_i(\underline{x}) F(\underline{x})~.
\end{split}
\end{equation}
Assuming that the complex parameter $q$ satisfies $|q|<1$, the contour $C_i/q$ will encircle $C_i$ as shown in figure \ref{contour}.
So, in order for \eqref{general_model} to hold, we must be able to shrink the contour $C_i/q$
back to $C_i$ so that we can take the sum over $i$ inside again.

We consider the implications of this condition in the following subsections.



\subsection{From full derivative to shift operators in times} \label{sec:derivation-of-constraints}
As we now have ensured that the insertions we consider are well-defined and vanishing, we will
proceed with rewriting them as some difference operators acting on times $\{t_k\}$,
in complete analogy with the non-$(q,t)$-deformed case.

\subsubsection{The generic constraint} \label{sec:the-generic-case}

The integrand of the left hand side of \eqref{eq:contour-shrink}
can be rewritten as (see Appendix~\ref{startingrelation})
\begin{align} \label{eq:startingrelation}
  \sum\limits_{i = 1}^{N} \ x_i^{\dagger} G_i(\underline{x}) F(\underline{x}) = \dfrac{F(\underline{x})}{(t-1)}
  \sum_{i=1}^N \Res_{\omega=q/x_i} \dfrac{d \omega}{\omega - z} \frac{c_q(1/\omega)}{c_q(q/\omega)} \prod\limits_{j = 1}^{N} \dfrac{q - t \omega x_j}{q - \omega x_j}  \prod\limits_{k = 1}^{M} \dfrac{\omega - t y_k}{\omega - y_k}~,
\end{align}
After we integrate over the eigenvalues, the left hand side of \eqref{eq:startingrelation}
can be rewritten as
\begin{equation} \label{nopoles}
\begin{split}
\oint_{C_1}  \dots \oint_{C_i}\dots \oint_{C_N} \prod_{j=1}^N \frac{d x_j}{x_j}\sum_{i=1}^N x_i^\dagger G_i(\underline{x}) F(\underline{x}) = \oint_{C_1} \dots \oint_{C_i}\dots \oint_{C_N} \prod_{j=1}^N \frac{d x_j}{x_j} \sum_{i=1}^NG_i(\underline{x}) F(\underline{x})
\end{split}
\end{equation}
provided we can shrink contour $C_i/q$ back to $C_i$. We can do this if there are no poles of the subintegral expression
in the shaded region. Let's look at different parts of the subintegral expression in detail.

As we said before, we treat the first part of $G_i(\underline{x})$, $\frac{1}{1-zx_i}$, as a formal power series
in non-negative powers of $z$, therefore, it doesn't contribute any poles.
The second part of $G_i(\underline{x})$, $\prod_{j \neq i} \frac{x_j-tx_i}{x_j-x_i}$ ,
can possibly have poles at $x_j = x_i$ for $j \neq i$ which could appear in the shaded area. However, the $q$-Pochhammer in the numerator of $F(\underline{x})$, $\prod_{k \neq l}(x_k/x_l;q)_\infty $ will precisely cancel these poles, and so there is no additional contribution from $G_i(\underline{x})$ as we shrink the contour.

Then let's consider $F(\underline{x})$. The first part, $\prod_{k \neq l} \frac{(x_k/x_l;q)_\infty }{(tx_k/x_l;q)_\infty}$,
can have poles for non-integer $\beta$. However, these poles are canceled by the numerator of $G_i(\underline{x})$.
In the second part of $F(\underline{x})$, the function $c_q(x_i)$ can, in principle, have poles in the shaded region.
One therefore needs to show that this doesn't happen for the model one is interested in,
otherwise equation \eqref{general_model} would acquire a non-trivial right hand side.

For the $(q,t)$-Gaussian model, the function $c_q(x_i)$ is given by \eqref{cqGaussian} and does not pick up any poles as we shrink the contour from encircling $(-\nu/q, \ \nu/q)$ to encircling $(-\nu, \nu)$: indeed, the only possible pole in this area is the point $x = \nu/q$ (a pole of $f_{\lambda}$) but it coincides with a zero of the $q$-Gaussian measure $(x^2 q^2 \nu^{-2}; q^2)_{\infty}$ and therefore is just a regular point.

For the gauge theory on $D^2 \times S^1$ model, the argument is analogous: for a given fugacity $m_f$ as we shrink the contour from encircling $(m_f^{-1} / q, \infty)$ to encircling $(m_f^{-1}, \infty)$, we do not encounter any poles, provided other fugacities $m_{f^{\prime}}$ are in generic positions.

Having demonstrated that for these two models there are no contributions from poles in the shaded area, from now on we proceed assuming this is the case, so that
\begin{align} \notag
\oint \prod_{j=1}^N \frac{dx_j}{x_j} \sum\limits_{i = 1}^{N}x_i^{\dagger} G_i(\underline{x}) F(\underline{x}) = \oint \prod_{j=1}^N \frac{dx_j}{x_j} \sum\limits_{i = 1}^{N} G_i(\underline{x}) F(\underline{x})~.
\end{align}
We then use the following algebraic identity (verified in Appendix~\ref{algebraicidentity})
\begin{equation}
\sum\limits_{i = 1}^{N} G_i(\underline{x}) = \sum\limits_{i = 1}^{N} \ \dfrac{1}{1 - z x_i} \ \prod\limits_{j \neq i} \dfrac{x_j - t x_i}{x_j - x_i} = \frac{1}{1-t} - \frac{1}{1-t} \prod_{i = 1}^{N} \frac{t - z x_i}{1 - z x_i}
\end{equation}
to obtain that
\begin{equation}
\begin{split}
\oint \prod_{j=1}^N \frac{dx_j}{x_j} \sum\limits_{i = 1}^{N} G_i(\underline{x}) F(\underline{x}) = \oint \prod_{j=1}^N \frac{dx_j}{x_j} & \bigg \{ \frac{1}{1-t} - \frac{1}{1-t} \prod_{i = 1}^{N} \frac{t - z x_i}{1 - z x_i}\bigg \}F(\underline{x})~.
\end{split}
\end{equation}

Let's now move to the right hand side of equation \eqref{eq:startingrelation}
where we also take an integral over eigenvalues. We want to interchange the integral over eigenvalues
with operation of taking the residue. But we cannot do this, since the points we take residues at
manifestly depend on the eigenvalues. Therefore, we change the sum over residues at $\omega=q/x_i$
for the sum of residues at $\omega = 0$ and $\omega=\infty$ (details are in Appendix~\ref{changecontour}).
Then we can safely exchange integral over eigenvalues with the residue operator.
\begin{equation} \label{eq:contour-transformation}
\begin{split}
\frac{1}{(t-1)} &\oint \prod_{j=1}^N \frac{dx_j}{x_j}   F(\underline{x}) \sum_{i=1}^N \Res_{\omega=q/x_i} \dfrac{d \omega}{\omega - z} \frac{c_q(1/\omega)}{c_q(q/\omega)} \prod\limits_{k = 1}^{N} \dfrac{q - t \omega x_k}{q - \omega x_k}  \prod\limits_{l = 1}^{M} \dfrac{\omega - t y_l}{\omega - y_l}\\
& \emph{} \hspace{-8ex} = \left(-\Res_{\omega = \infty} -\frac{1}{(t-1)} \Res_{\omega=0}\right) \dfrac{d \omega}{\omega - z} \frac{c_q(1/\omega)}{c_q(q/\omega)} \prod\limits_{l = 1}^{M} \dfrac{\omega - t y_l}{\omega - y_l} \oint \prod_{j=1}^N \frac{dx_j}{x_j}   F(\underline{x}) \prod\limits_{k = 1}^{N} \dfrac{q - t \omega x_k}{q - \omega x_k}
\end{split}
\end{equation}
We can only do the above transformation provided that the subintegral expression has no other poles except $q/x_i$, $0$ and $\infty$.
If there would be extra poles they would contribute to the right hand side of the Ward identities.
Then Ward identities would no longer be $q$-Virasoro constraints, at least not in the representation \eqref{eq:heisenberg-representation}. So, we have an additional requirement that $\frac{c_q(1/\omega)}{c_q(q/\omega)}$ should have no poles
except at $0$ and $\infty$.
This requirement should be checked for every concrete model one is interested in.
For instance, for the $(q,t)$-Gaussian matrix model it is straightforward to see that this is, indeed, the case.
For the gauge theory on $D^2 \times S^1$, however, the fugacity for the fundamental, $m_f$, that appears
in the denominator of \eqref{blockIntegral} could cause such poles.
Therefore, the dependence on the fundamental fugacities can be studied in the framework of $q$-Virasoro constraints
only as a formal power series.
Luckily, the dependence on the anti-fundamental fugacities $\bar{m}_f$ need not be understood formally,
one can study it non-perturbatively.

Hence, equation \eqref{eq:startingrelation} becomes (after some massaging that involves calculating residue at $\omega=\infty$)
\begin{equation}
\begin{split}
\Bigg [\oint \prod_{j=1}^N \frac{dx_j}{x_j} & \bigg \{ \frac{1}{1-t} - \frac{1}{1-t} \prod_{i = 1}^{N} \frac{t - z x_i}{1 - z x_i}\bigg \}F(\underline{x}) - \frac{t^N}{(t-1)} \left. \frac{c_q(z)}{c_q(qz )} \right|_{z =0} \oint \prod_{i=1}^N \frac{dx_i}{x_i}  F(\underline{x})\\
& \emph{} \hspace{-13ex} +\frac{1}{(t-1)} \oint_{\omega=0} \dfrac{d \omega}{\omega - z} \frac{c_q(1/\omega)}{c_q(q/\omega)} \prod\limits_{i = 1}^{M} \dfrac{\omega - t y_i}{\omega - y_i} \oint \prod_{j=1}^N \frac{dx_j}{x_j} F(\underline{x}) \prod\limits_{k = 1}^{N} \dfrac{q - t \omega x_k}{q - \omega x_k}\Bigg] \ = \ 0~.
\end{split}
\end{equation}
Using the explicit form of the generating function given in \eqref{generatingfn} and the simple fact that $c_q(z) / c_q(qz )\big|_{z =0} = Q^{-1} p^{-1} q^{-\sqrt{\beta} \alpha}$ where $\alpha$ is the momentum parameter from \eqref{cqgeneral}, we can write the above in terms of the partition function $\mathcal{Z}$ using shifts in the time-variables $\{t_k\}$:
\begin{equation} \label{firstconstraint}
\boxed{ \ \ \ \ \
\begin{split}
 & -\Bigg(p^{-1} q^{-\sqrt{\beta} \alpha} + 1\Bigg) \ \mathcal{Z}\big\{t_k\big\}+t^N \mathcal{Z}\left\{t_k \rightarrow t_k+\frac{z^k}{k}(1-t^{-k} ) \right\}  \\
& +\oint_{\omega=0}  \frac{d \omega}{\omega - z} \ \frac{c_q(1/\omega)}{c_q(q/\omega)} \ e^{\sum_{k=1}^\infty \frac{(1-q^k)t_k}{\omega ^k} } \ \mathcal{Z}\left\{t_k \rightarrow t_k+\frac{\omega^k(1-t^k)}{kq^k} \right\} = 0~,
\end{split}
\ \ \ \ \ }
\end{equation}
which is then our first constraint equation.

Here we have rewritten certain products as exponents of logarithms, for instance
\begin{align}
  \prod_{i=1}^M \frac{\omega - t y_i}{\omega - y_i}
  = \exp \left ( \sum_{k=1}^\infty \frac{(1-t^k)}{\omega^k} \frac{\left(\sum_{i=1}^M y_i^k \right)}{k} \right )
  = \exp \left ( \sum_{k=1}^\infty \frac{(1-q^k)}{\omega^k} t_k \right );
  \ \ \ t_k = \frac{(1-q^k)}{(1-t^k) k} \sum_{i=1}^M y_i^k~.
\end{align}
Let us proceed to the second constraint.

\subsubsection{The special additional constraint} \label{sec:special-additional-case}

There is another insertion of full $q$-derivative that one can consider,
that leads to an additional equation for the partition function.
In the case of $(q,t)$-deformed matrix model, this equation is necessary to make the system of equations
closed and to be able to find all the correlators. In the classical limit this equation goes into the string equation.
To obtain it, we begin with the relation (verified in Appendix~\ref{startingrelation})
\begin{equation}
  \sum\limits_{i = 1}^{N} \ x_i^{\dagger} \tilde{G}_i(\underline{x}) F(\underline{x}) = \dfrac{F(\underline{x})}{(t-1)}
 \sum_{i=1}^N \Res_{\omega=q/x_i} d \omega \frac{c_q(1/\omega)}{c_q(q/\omega)} \prod\limits_{j = 1}^{N} \dfrac{q - t \omega x_j}{q - \omega x_j}  \prod\limits_{k = 1}^{M} \dfrac{\omega - t y_k}{\omega - y_k},
\end{equation}
where now
\begin{equation}
\tilde{G}_i(\underline{x}) = \dfrac{1}{x_i} \ \prod\limits_{j \neq i} \dfrac{x_j - t x_i}{x_j - x_i}~.
\end{equation}
Proceeding similarly to above, the first step is to take a contour integral of the left hand side and use the algebraic identity
\begin{equation}
\sum\limits_{i = 1}^{N} \tilde{G}_i(\underline{x}) = \sum\limits_{i = 1}^{N} \ \dfrac{1}{ x_i} \ \prod\limits_{j \neq i} \dfrac{x_j - t x_i}{x_j - x_i} = \sum\limits_{i = 1}^{N} \dfrac{1}{ x_i}
\end{equation}
to get
\begin{equation}
\oint \prod_{j=1}^N \frac{dx_j}{x_j} \sum\limits_{i = 1}^{N} x_i^\dagger \tilde{G}_i(\underline{x}) F(\underline{x}) = \oint \prod_{j=1}^N \frac{dx_j}{x_j}  \bigg \{ \sum\limits_{i = 1}^{N} \dfrac{1}{ x_i} \bigg \}F(\underline{x})~.
\end{equation}

We then transform the right hand side by changing the residues from $\omega=q/x_i$ to $0$ and $\infty$
and swapping integration over eigenvalues with taking the residue.
Writing the result using $\mathcal{Z}\{t_k\}$ we get after some calculation
\begin{equation}
\boxed{
\ \ \ \ \ \begin{split} \label{secondconstraint}
\oint & \prod_{j=1}^N \frac{dx_j}{x_j} \bigg \{ \sum\limits_{i = 1}^{N} \dfrac{1}{ x_i} \bigg \}F(\underline{x}) \Bigg(1 - q^{-\sqrt{\beta} \alpha} \Bigg) + q^{-\sqrt{\beta} \alpha} \ \mathcal{Z}\big\{t_k\big\} t_1 \frac{(1-q)}{(1-t)} \\
&+\frac{1}{(t-1)} \Res_{\omega=0} d \omega \frac{c_q(1/\omega)}{c_q(q/\omega)} e^{\sum_{k=1}^\infty \frac{(1-q^k)t_k}{ \omega^k} } \mathcal{Z}\left\{t_k \rightarrow t_k +\frac{\omega^k(1-t^k)}{kq^k}\right\} =0~.
\end{split} \ \ \ \ \
}
\end{equation}
Note that the complicated first term vanishes in the case $\alpha = 0$; this will be useful later.
Otherwise, it is not clear what to do with it: it is an average of $\langle \text{tr} X^{-1} \rangle$ and as such
it is not expressible as any differential operator in $\{t_k\}$ acting on $\mathcal{Z}$. Perhaps, the set of times
could be enlarged to be able to also generate correlators of traces of negative powers of $X$, however,
investigation of this possibility is beyond the scope of the present paper.

\subsubsection{Combining generic and special constraints}

We then want to combine our two constraint equations. The key point here is the observation that the first constraint in equation \eqref{firstconstraint} can be written as
\begin{equation}
\begin{split}
& \Bigg (p^{-1} q^{-\sqrt{\beta} \alpha} + 1\Bigg) \ \mathcal{Z}\big\{t_k\big\} - t^N \mathcal{Z}\left\{t_k \rightarrow t_k+\frac{z^k}{k}(1-t^{-k} ) \right\}  \\
& = \frac{c_q(1/z)}{c_q(q/z)} \ e^{\sum_{k=1}^\infty \frac{(1-q^k)t_k}{z ^k} } \ \mathcal{Z}\left\{t_k \rightarrow t_k+\frac{z^k(1-t^k)}{kq^k} \right\} + f(1/z)
\end{split}
\end{equation}
for some function $f(1/z)= \sum_{m=1}^\infty \frac{d_m}{z^m}$ containing only negative powers in $z$. The second constraint equation in \eqref{secondconstraint} then gives us the coefficient of $\frac{1}{z}$ in this series,
allowing to combine into
\begin{equation}
\hspace{-3ex} \boxed{
\ \ \
\begin{split}
t^N &\mathcal{Z}\left\{t_k \rightarrow t_k+\frac{z^k}{k}(1-t^{-k} ) \right\} \ + \ \frac{c_q(1/z)}{c_q(q/z)} \  e^{\sum_{k=1}^\infty \frac{(1-q^k)t_k}{z ^k} } \ \mathcal{Z}\left\{t_k \rightarrow t_k+\frac{z^k(1-t^k)}{kq^k} \right\}\\
& = \big( p^{-1} q^{-\sqrt{\beta} \alpha} + 1 \big) \ \mathcal{Z}\big\{t_k\big\} + \frac{1-t}{z} \oint \prod_{j=1}^N \frac{dx_j}{x_j} \bigg \{ \sum\limits_{i = 1}^{N} \dfrac{1}{ x_i} \bigg \}F(\underline{x}) \Bigg\{1-q^{-\sqrt{\beta} \alpha} \Bigg\} \\
& + \frac{1-q}{z} \ p^{-1} q^{-\sqrt{\beta} \alpha} \ t_1 \mathcal{Z}\big\{t_k\big\} + \sum_{m=2}^\infty \frac{d_m}{z^m}
\end{split} \label{eq18}
\ \ \ \ \
}
\end{equation}
which the generating function $\mathcal{Z}\{t_k\}$ must satisfy.
\\
\\
As an example and for comparison with earlier work on $q$-Virasoro, let us consider the special case with no potential, i.e. $N_f = 0$ and $\alpha = 0$. It can be seen that in this case the above constraint takes a simple form
\begin{equation} \label{eq:no-potential-virasoro}
\begin{split}
t^N \mathcal{Z}\{t_k \rightarrow t_k+\frac{z^k}{k}(1-t^{-k}) \} + & q^{-1}t^{1-N}  e^{\sum_{k=1}^\infty \frac{(1-q^k)t_k}{z^k} } \mathcal{Z}\{t_k \rightarrow t_k+\frac{z^k(1-t^k)}{kq^k} \} \\
= & (p^{-1}+1)\mathcal{Z}\{t_k\} +\frac{tq^{-1}(1-q)t_1\mathcal{Z}\{t_k\} }{z} + \sum_{m=2}^\infty \frac{c_m}{z^m}~,
\end{split}
\end{equation}
for some coefficients $c_m$. Comparing this to the construction in \cite{Shiraishi:1995rp} and the $q$-Virasoro constraint in equation  \eqref{qVirasoroConstraint} (with $z \rightarrow 1/z$) we are considering
\begin{equation}
\psi(1/z) T(1/z) \mathcal{Z}\{t_k\} = c_0 + \frac{c_1}{z} +\sum_{m=2}^\infty \frac{c_m}{z^m}
\end{equation} 
with $c_0$ and $c_1$ being the coefficients of the $z^0$ and $z^{-1}$ terms respectively, and 
\begin{equation}
\psi (1/z) = p^{-\frac{1}{2}}\exp \left \{ \sum_{k>0}^\infty \frac{(1-q^k)t_k}{(1+p^k)z^k}  \right \}~.
\end{equation}
To then extract the action of $T_0$ and $T_{-1}$ on $\mathcal{Z}\{t_k\}$ we expand the stress tensor current $T(1/z)=\sum_{n \in \mathbb{Z}}T_n z^{n}$ and compute
\begin{equation}
\begin{split}
T(1/z) \mathcal{Z}\{t_k\} = & \psi (1/z)^{-1} \left( (p^{-1}+1)\mathcal{Z}\{t_k\} +\frac{tq^{-1}(1-q)t_1\mathcal{Z}\{t_k\} }{z} + \sum_{m=2}^\infty \frac{c_m}{z^m} \right ) \\
= & \left( p^{\frac{1}{2}}+p^{-\frac{1}{2}} \right ) \mathcal{Z}\{t_k\}+ \psi (1/z)^{-1} \left(  \sum_{m=2}^\infty \frac{c_m}{z^m} \right )
\end{split}
\end{equation}
from which we can identify the eigenvalue equations
\begin{equation}
\begin{split}
T_0 \mathcal{Z}\{t_k \} = & \left( p^{\frac{1}{2}}+p^{-\frac{1}{2}} \right) \mathcal{Z}\{t_k \} \\
T_{-1} \mathcal{Z}\{t_k \} = & 0~.
\end{split}
\end{equation}
This can also be obtained using the free field representation of \cite{Nedelin:2016gwu} as shown in appendix \ref{FreeField}, and so these two viewpoints are equivalent.

\section{Recursive solution} \label{recursive}

{
  The aim of this section is to show how $q$-Virasoro constraints
  can be used to recursively determine all normalized correlators
  of a given eigenvalue model. We will use the $(q,t)$-Gaussian matrix model
  as an example, but models with more involved potentials can be treated
  similarly -- they just require more initial conditions to start the recursion.
  This situation is completely analogous to the usual, non-$(q,t)$-deformed case,
  where one needs to choose a set of integration contours (i.e. the ``phase'' of the model)
  to define its correlators \cite{Cordova:2016jlu}.
}
\\
\\
{
  The $q$-Virasoro constraints \eqref{eq18}, specialized for $(q,t)$-Gaussian matrix model, read
  \begin{align} \label{eq:q-vir-gauss-combined}
    &
    t Q^{-1} \left ( 1 - q^2(1-q) y^2 \right ) \exp \left ( \sum_{k=1}^\infty (1-q^k) t_k y^k \right )
    \mathcal{Z} \left \{ t_k \rightarrow t_k + \frac{(1-t^k)}{k q^k}\frac{1}{y^k} \right\}
    \\ \notag &
    + q Q \mathcal{Z} \left \{ t_k \rightarrow t_k + \frac{(1-t^{-k})}{k} \frac{1}{y^k} \right\}
    - (q + t) \mathcal{Z} \left \{ t_k \right \} - y t (1-q) t_1 \mathcal{Z}\left \{ t_k \right \} = \sum_{m=0}^\infty c_m y^{m+2}
  \end{align}
where $Q=t^N$ and $c_m$ are coefficients of the expansion in $y$. We search for a formal power series solution to this equation
  \begin{align}
    \mathcal{Z}\{t_k \} = \sum_{n=0}^\infty \frac{1}{n!} \sum_{d=0}^\infty \sum_{l_1 +\dots +l_n = d} C_{l_1 \dots l_n} t_{l_1} \dots t_{l_n}
    = \sum_{d=0}^\infty \mathcal{Z}_d\{t_k \},
  \end{align}
where $C_{l_1 \dots l_n}$ are correlators. $\mathcal{Z}_d\{t_k \}$ is defined to be the degree $d$ component of the partition function with respect to the action of the operator $\sum_{d=1}^\infty d\; t_d \frac{\partial}{\partial t_d}$.
\\
\\
Extracting the coefficient in front of $y^{-m}, m = -1, 0, 1, ...$
  and of particular degree $d$ in times $t_n$ from equation~\eqref{eq:q-vir-gauss-combined} we get
  \begin{align}\label{eq:q-vir-gauss-md}
    & t Q^{-1} q^2 (1-q) s_{m+2} \left ( \left \{p_k = \frac{(1-t^k)}{q^k} \frac{\partial}{\partial t_k} \right \} \right ) \mathcal{Z}_{d+2}\{t_k \}
    \\ \notag
    & = t Q^{-1} s_{m} \left ( \left \{p_k = \frac{(1-t^k)}{q^k} \frac{\partial}{\partial t_k} \right \}\right ) \mathcal{Z}_{d}\{t_k \}
    \\ \notag &
    \quad + q Q \sum_{p=0}^{d - m} s_p\left ( \left\{p_k = -(1-q^k) k t_k \right \}\right )
    s_{p+m}\left ( \left\{ p_k = (1-t^{-k}) \frac{\partial}{\partial t_k} \right \} \right ) \mathcal{Z}_{d}\{t_k \}
    \\ \notag & \quad - \delta_{m,0} (q+t) \mathcal{Z}_{d}\{t_k \}
    + \delta_{m,-1} q (1 - q) t_1 \mathcal{Z}_{d}\{t_k \},
  \end{align}
  where $s_m(p_1,\dots,p_m)$ is Schur polynomial for symmetric representation $[m]$, see \eqref{eq:symm-schur-poly}.

By considering coefficients in front of all monomials in times $t_n$ in equations \eqref{eq:q-vir-gauss-md},
we obtain an overdetermined system of linear equations for the correlators $C_{l_1 \dots l_n}$.
  From this system one can actually pick certain equations in a clever way in specific order and obtain a recursive formula
  for the correlators. Suppose we wish to find the correlator $C_{\lambda_1 ... \lambda_{\bullet - 1} \lambda_{\bullet}}$. Here $\lambda_{\bullet}$ means the last part of the partition $\lambda$ and we assume that indices of the correlator are sorted in descending order such that $\lambda_{\bullet}$ is the smallest index (the correlator is of course symmetric as a function of its indices).
  We then consider equation \eqref{eq:q-vir-gauss-md} with $m + 2 = \lambda_\bullet, \ \ d + 2 = |\lambda|$
  and apply $\Bigg{|}_{\vec{t} = 0} \frac{\partial^{\bullet - 1}}{\partial t_{\lambda_1} \dots \partial t_{\lambda_{\bullet - 1}}}$
  to it (i.e. setting all times to zero after taking the partial derivative). We then get
  \begin{align} \label{eq:q-vir-correlators-explicit}
    t Q^{-1} q^2(1-q) \frac{1}{\lambda_\bullet} & \frac{(1 - t^{\lambda_\bullet})}{q^{\lambda_\bullet}}
    C_{\lambda_1 \dots \lambda_\bullet}  =
    - t Q^{-1} q^2(1-q) \sum_{\substack{|\vec{\mu}| = \lambda_\bullet \\ l(\vec{\mu}) \geq 2}}
    \frac{1}{l(\vec{\mu})!} \left ( \prod_{a \in \vec{\mu}} \frac{(1 - t^a)}{q^a a} \right )
    C_{\lambda_1 \dots \lambda_{\bullet - 1} \mu_1 \dots \mu_\bullet}
    \\ \notag &
    - \delta_{\lambda_\bullet, 2} (q + t) C_{\lambda_1 \dots \lambda_{\bullet - 1}}
    + \delta_{\lambda_\bullet, 1} ((\#_\lambda 1) - 1) q (1 - q) C_{\lambda_1 \dots \lambda_{\bullet - 2}}
    \\ \notag &
    + t Q^{-1} \sum_{|\vec{\mu}| = \lambda_\bullet - 2}
    \frac{1}{l(\vec{\mu})!} \left ( \prod_{a \in \vec{\mu}} \frac{(1 - t^a)}{q^a a} \right )
    C_{\lambda_1 \dots \lambda_{\bullet - 1} \mu_1 \dots \mu_\bullet}
    \\ \notag &
    + q Q \sum_{\nu \subseteq \lambda \setminus \lambda_\bullet}
    \left ( \prod_{a \in \nu} (-1) (1 - q^a) \right ) \sum_{|\vec{\mu}| = |\nu| + \lambda_\bullet - 2}
    \frac{1}{l(\vec{\mu})!} \left ( \prod_{a \in \vec{\mu}} \frac{(1 - t^{-a})}{a} \right )
    C_{\lambda \setminus \{\lambda_\bullet, \nu\} \vec{\mu}}
  \end{align}
  Here $\vec{\mu}$ denotes an \textit{ordered} sequence of positive integer numbers (it can be an empty sequence), as opposed to a partition which is \textit{sorted} sequence of positive integer numbers and  $l(\vec{\mu})$ is the length of the sequence $\vec{\mu}$.
  $\#_\lambda 1$ is the number of parts of $\lambda$ equal to 1, $\lambda \setminus \lambda_\bullet$ is partition $\lambda$
  without part $\lambda_\bullet$ and $\lambda \setminus \{\lambda_\bullet, \nu\}$ is partition $\lambda$ with part $\lambda_\bullet$
  and all parts of $\nu$ removed.
  
  The formula \eqref{eq:q-vir-correlators-explicit} is indeed a recursion:
  correlators on all lines of the right hand side except the first one have sum of indices equal to $|\lambda| - 2$;
  the correlators on the first line, though having sum of indices equal to $|\lambda|$, all have a minimal
  index that is strictly less than $\lambda_\bullet$ (thanks to the condition $l(\vec{\mu}) \geq 2$).

  Supplemented with the initial conditions $C_\emptyset = 1$ and $C_1 = 0$, this recursion allows us to calculate any particular correlator in a finite number of steps.

  \paragraph{Example:}The first step of the recursion looks like
  \begin{align}
    t Q^{-1} q^2(1-q) \frac{(1-t)}{q} C_{1,1} & = q (1-q) + q Q (-1) (1-q)
    \\ \notag & \Longrightarrow
    C_{1,1} = \frac{Q(1-Q)}{t(1-t)}
    \\ \notag
    t Q^{-1} q^2(1-q) \frac{1}{2} \frac{(1-t^2)}{q^2} C_2 & = - t Q^{-1} q^2 (1-q) \frac{1}{2} \frac{(1-t)^2}{q^2} C_{1,1}
    - (q + t) + t Q^{-1} \frac{1}{0!} + q Q
    \\ \notag & \Longrightarrow
    C_2 = \frac{(1-Q)(2 t - Q - q Q + t Q - q t Q)}{t(1-q)(1-t^2)}
  \end{align}

}

{
  In the case of non-$(q,t)$-deformed Hermitean matrix model, the Virasoro constraints
  can be used to derive the so-called $\hat{W}$-operator representation of the partition function
  \cite{Morozov:2009xk}. This derivation can be easily generalized to the case of Gaussian $\beta$-ensemble.
  However, at the moment we do not know how to generalize it to the $(q,t)$-Gaussian matrix model,
  as well as more complicated $(q,t)$-models. Existence of $\hat{W}$-operator form (the cut-and-join equation)
  is intimately linked with the existence of the spectral curve topological recursion
  for the model \cite{Eynard:2007kz,Alexandrov:2006qx,Alexandrov:2009gn,Orantin:2008hq} and often signifies connections with enumerative geometry \cite{doi:10.1112/jlms/jdv047}. Pursuing these directions is definitely worth it in the future.
}
\section{Gauge theory applications} \label{gaugetheory}
{
From the point of view of gauge theories the normalized correlators $C_{l_1 \dots l_n}$ are not very interesting.
Instead, of primary interest are averages of Schur functions, which on the gauge theory
side correspond to supersymmetric Wilson loops (see \cite{Nedelin:2016gwu} and references therein).
In the gauge theory on $D^2 \times S^1$ 
 the expectation value of the Wilson loop along $S^1$  in representation $\lambda$  corresponds to the following expression 
\begin{equation}
\left\langle s_\lambda \right \rangle  = \oint_c \prod_{i=1}^N \frac{dx_i}{x_i} \prod_{k \neq l=1}^N \frac{(x_k/x_l;q)_\infty }{(tx_k/x_l;q)_\infty} \ \prod_{j=1}^N \prod_{f=1}^{N_f} \frac{(qx_j\overline{m}_f;q)_\infty }{(x_jm_f;q)_\infty } \  s_\lambda (x)~,
\end{equation}
 where $s_\lambda$ is Schur polynomial for the representation $\lambda$. 
 If we choose $N_f=2$ with two opposite anti-fundamental masses $\pm \overline{m}$ having the special values $\pm \overline{m}_f=\pm q(1-q)^{\frac{1}{2}}$
 and  without fundamental contributions
 we can use the previously derived results.}
This choice of masses is by no means distinguished from the gauge theory point of view. Moreover, one could easily
insert arbitrary values of the two anti-fundamental masses on the matrix model side as well,
and this will not complicate $q$-Virasoro constraints at all
(the factor $\left( 1 - q^2(1-q) y^2 \right)$ in the first term of the equation \eqref{eq:q-vir-gauss-combined}
will be replaced by $\left( 1 - \overline{m}_1 y \right) \left( 1 - \overline{m}_2 y \right)$
but it will still remain \textit{quadratic}, so non-Gaussian issues will still not arise).
In what follows we present formulas at special values of the masses for conciseness.

Reexpanding the partition function in the basis of Schur functions (using Cauchy identity)
\begin{equation}
  \mathcal{Z}\{t_k\} = \sum_{n=0}^\infty \frac{1}{n!} \sum_{l_1,\dots,l_n} C_{l_1 \dots l_n} t_{l_1} \dots t_{l_n}
  = \left \langle \exp \left ( \sum_{k=1}^\infty \text{tr} \Phi^k t_k \right ) \right \rangle
  = \sum_\lambda s_\lambda (\{ p_k = k t_k \}) \left \langle s_\lambda ( \{p_k = \text{tr}\Phi^k \}) \right \rangle
\end{equation}
it is straightforward to obtain constraints on Schur averages. In the case of the $(q,t)$-Gaussian model, the first few of them are:
\begin{align} \label{eq:vir-const-schur}
  \left\langle s_2 \right \rangle =  \left\langle s_\emptyset \right \rangle \Bigg( & Q^2
    \left(\frac{-5 q-3}{2 (q-1) t}+\frac{5 q-1}{2 (q-1) (t-1)}+\frac{2}{(q-1)
      (t+1)}\right)
    \\ \notag &
    +Q \left(\frac{-5 q-3}{2 (q-1) (t-1)}+\frac{5 q+3}{2 (q-1)
      t}\right)+\frac{2}{(q-1) (t-1)}-\frac{2}{(q-1)
      (t+1)} \Bigg)
    \\ \notag
    \left\langle s_{1,1} \right \rangle = \left\langle s_\emptyset \right \rangle
    \Bigg( & Q^2 \left(\frac{-3 q-1}{2 (q-1) (t-1)}+\frac{3 q+5}{2 (q-1)
      t}-\frac{2}{(q-1) (t+1)}\right)
    \\ \notag & +Q \left(\frac{-3 q-5}{2 (q-1) t}+\frac{3
    q+5}{2 (q-1) (t-1)}\right)-\frac{2}{(q-1) (t-1)}+\frac{2}{(q-1)
    (t+1)} \Bigg)
\end{align}

\begin{align}
  \notag
  \left\langle s_4 \right \rangle = \left\langle s_2 \right \rangle & \Bigg( \Big(\frac{-59 q-21}{8 (q-1)
    t}+\frac{7 q-3}{8 (q-1) (t-1)}+\frac{8 q+9}{2 (q-1) (t+1)}+\frac{-6 q^2+5
    t q-5 q-3 t-5}{2 (q-1) \left(t^2+1\right)}
  \\ \notag & \quad \quad +\frac{1205 q^3+801 q^2+855
    q+211}{192 (q-1) t^2}+\frac{-241 q^3-117 q^2-27 q+1}{64 (q-1) t^3}\Big)
  Q^2
  \\ \notag & \quad +\left(\frac{-7 q-1}{8 (q-1) (t-1)}-\frac{21}{8
    t}+\frac{4}{t+1}+\frac{-t q+5 q+3 t-7}{2 (q-1) \left(t^2+1\right)}\right)
  Q
  \\ \notag & \quad +\frac{1}{2 (q-1) (t-1)}-\frac{17}{2 (q-1) (t+1)}+\frac{8 t-1}{(q-1)
      \left(t^2+1\right)}\Bigg)
  \\ \notag + \left\langle s_{1,1} \right \rangle
  & \Bigg(\Big(\frac{1-5 q}{8 (q-1) (t-1)}-\frac{3 (13 q+3)}{8 (q-1) t}+\frac{8 q+7}{2 (q-1)
    (t+1)}+\frac{-10 q^2+3 t q-3 q-5 t-3}{2 (q-1)
    \left(t^2+1\right)}
  \\ \notag & \quad \quad +\frac{241 q^3+445 q^2+267 q+71}{64 (q-1) t^2}-\frac{5
    \left(241 q^3+117 q^2+27 q-1\right)}{192 (q-1) t^3}\Big)
  Q^2
  \\ \notag & \quad +\left(\frac{5 q+3}{8 (q-1) (t-1)}-\frac{9}{8 t}+\frac{4}{t+1}+\frac{-7
    t q+3 q+5 t-1}{2 (q-1) \left(t^2+1\right)}\right) Q
  \\ \notag & -\frac{1}{2 (q-1)
    (t-1)}-\frac{15}{2 (q-1) (t+1)}+\frac{8 t+1}{(q-1)
    \left(t^2+1\right)}\Bigg)
\end{align}
 These expressions are examples of the expectation values of Wilson loops in $D^2 \times S^1$ with the specific choices for the matter sector. 
  In principle one may obtain the formulas for general values of fundamental mass by using the shifts in times. 
At the moment we do not know of a more direct way to write down the expessions of the form \eqref{eq:vir-const-schur}. Moreover, in the case 
of usual, non-$(q,t)$-deformed matrix models the Virasoro constraints look rather cumbersome in the Schur basis \cite{Adler2003} so there
 is no reason to expect them to be simpler for the $(q,t)$-deformation.

It is curious that from the point of view of $(q,t)$-deformed matrix models
it is natural to study averages not of Schur functions, but rather of Macdonald functions ${\cal M}_\lambda (\{p_k\})$ \cite{Morozov:2018eiq}.
Computer experiments show that in many examples Macdonald averages have a very simple, explicit form.
For instance, for $(q,t)$-Gaussian matrix model one has
\begin{align}
  \left \langle {\cal M}_\lambda \right \rangle = {\cal M}_\lambda \left( \left\{p_k = (1 + (-1)^k) \frac{(1-q)^{k/2}}{(1-t^k)} \right \}\right )
  \prod_{(i,j)\in \lambda} \frac{1 - t^{N+1-i}q^{j-1}}{1-q}
\end{align}
though at the moment we do not know how to prove this formula.
But, at least in principle, Schur averages could be obtained from the vector of Macdonald averages
by applying the inverse matrix of $(q,t)$-Kostka coefficients \cite{qtKostka}.

The techniques suggested in this work can be combined with the free field realization of the gauge theory on squashed $S^3_{(\omega_1, \omega_2)}$
 \cite{Nedelin:2016gwu}. There exists the generating function for the expectation values of supersymmetric Wilson loops 
\begin{equation}
   \mathcal{Z}\{t_k, \tilde{t}_k\}  = \int_{i \mathbb{R}^N} \prod_{i=1}^N  dX_i \prod_{k \neq l=1}^N \frac{S_2 (X_k - X_l | \underline{\omega})}{S_2 (M_a + X_k -X_l | \underline{\omega})}  e^{\sum\limits_{j=1}^N V(X_j)}
      e^{[\sum\limits_{n=0}^\infty t_n \sum\limits_{j=1}^N e^{2\pi i n X_j /\omega_1} + \sum\limits_{n=0}^\infty \tilde{t}_n \sum\limits_{j=1}^N e^{2\pi i n X_j /\omega_2}]}~,
 \end{equation}
  here we use the notations from \cite{Nedelin:2016gwu}. Upon certain assumptions this generating function satisfies two commuting sets of $q$-Virasoro constraints. 
   It is straightforward to repeat the analysis from Section \ref{sec:q-diff-op} keeping in mind that there is a change of variables $x = e^{2\pi i X/\omega_1}$ (or $x = e^{2\pi i X/\omega_2}$
    for another copy of $q$-Virasoro). We leave this analysis for future studies.

\section{Conclusion} \label{conclusion}


In this paper we showed how to derive the set of Ward identities which a $(q,t)$-eigenvalue model satisfies, starting from an integral representation of it.
The derivation is based on a clever insertion of a full $q$-derivative under the integral. Under certain mild assumptions (which are to be checked in each particular case) these Ward identities are nothing but $q$-Virasoro constraints, which before could be derived only if the free field realization (in particular, the corresponding screening current) for the model was known.

The advantage of the method we present here is that it can be applied to any $(q,t)$-eigenvalue model
directly -- there is no need to know the free field representation beforehand.
Hence, the situation with $(q,t)$-eigenvalue models now becomes very similar to the situation
with $\beta$-ensembles, where Virasoro constraints can be derived both from free field realization
and from insertion of full derivatives under the integral.
We think that having these two complementary perspectives is useful, as each of them highlights different
aspects of the eigenvalue models.

It is worth noting that from the $q$-derivative insertion method one quite naturally obtains \textit{more} constraints
than from free field realization. Namely, there is an additional constraint $T_{-1} \mathcal{Z} = 0$.
During its derivation one has to demand that the coefficient in front of $\left \langle \text{tr} X^{-1} \right \rangle$
vanishes, so that one stays in the usual setting of matrix models. On the free field representation side this extra equation is satisfied if one restricts
attention to highest weights with zero momentum and it is very easy to overlook this special case.

There are several directions of further research that we think are interesting:

\begin{itemize}
\item \textbf{Application to concrete matrix models:} The $q$-Virasoro constraints \eqref{eq18}
  can be viewed as a definition of the matrix model.
  A particular integral representation is then just one instance (or ``phase'')
  that satisfies this definition.
  It would be very interesting to study alternative phases of several objects
  which are of crucial importance in mathematical physics and are known to satisfy $q$-Virasoro constraints,
  such as ABJ(M) model and Nekrasov functions.
\item \textbf{$\hat{W}$-operator representation:} In the case of usual Hermitean Gaussian matrix model the $\hat{W}$-operator
  representation is a consequence of Virasoro constraints. We hope that the explicit form of $q$-Virasoro constraints
  \eqref{eq:q-vir-correlators-explicit} will be useful in deriving $\hat{W}$-operator representation
  for the $(q,t)$-Gaussian model.
The $\hat{W}$-operator representation would then readily imply connection to enumerative geometry.
\item \textbf{Large $N$ limit and related questions:} For non-$(q,t)$-deformed matrix models the Virasoro constraints
  can be used to derive the so-called loop equations, which then may be solved in the large $N$ limit \cite{Eynard:2008we}.
  A particular solution, satisfying certain mild assumptions, takes the form of a certain universal recursion procedure,
  the \textit{spectral curve topological recursion}. Information about the concrete matrix model
  is translated into the initial algebrogeometric data for the recursion -- the spectral curve.
  The spectral curve topological recursion is related to the Givental formalism for Cohomological field theories.
  It would be very interesting to see what the $(q,t)$-spectral curve topological recursion and
  $(q,t)$-Givental formalism look like.
\item \textbf{Relation to Macdonald operators:} The $q$-difference operators that we insert under the integral
  look very similar to the celebrated Macdonald operators \cite[equation (31)]{Shiraishi:1995rp}.
  By understanding this connection better one could, perhaps, better understand the integrability
  of $(q,t)$-eigenvalue models.
\end{itemize}

We hope that these questions would inspire some great future research.

\bigskip
{\bf Acknowledgements:} We thank Fabrizio Nieri for  discussions.
 The work of RL, AP and MZ  is supported in part by Vetenskapsr\r{a}det
 under grant \#2014-5517, by the STINT grant, and by the grant  ``Geometry and Physics"  from the Knut and Alice Wallenberg foundation. AP is also partially supported by RFBR grants 16-01-00291, 18-31-20046 mol\_a\_ved and 19-01-00680 A.
 SS is partially supported by RFBR grant 18-31-20046 mol\_a\_ved.

\appendix
\section{Special functions}
A summary of the special functions used throughout the paper. The definition of the $q$-Pochhammer symbol is
\begin{equation} \label{qpochhammer}
(x;q)_\infty = e^{-\sum_{k>0}\frac{x^k}{k(1-q^k)}}=\prod_{k=0}^\infty (1-xq^k)
\end{equation}
valid for $|q|<1$, and the finite $q$-Pochhammer is given by
\begin{equation}
(x;q)_n = \frac{(x;q)_\infty}{(q^nx;q)_\infty}=\prod_{k=1}^{n-1}(1-xq^k).
\end{equation}
The $\theta $ function is defined by
\begin{equation} \label{thetafunction}
\theta (x;q)= (x;q)_\infty (qx^{-1};q)_\infty~.
\end{equation}
The Schur polynomial $s_m(p_1,\dots,p_m)$ in symmetric representation $[m]$ is given by
\begin{equation}
\begin{split}
    & \exp \left(\sum_{k=1}^\infty \frac{z^k p_k}{k}\right) = \sum_{m=0}^\infty z^m s_m(p_1,\dots,p_m)
    \\ \label{eq:symm-schur-poly}
    & s_m(p_1,\dots,p_m) = \prod_{\lambda \vdash m} \prod_{i=1}^{l(\lambda)} \frac{p_i}{i}
    \prod_{j=1}^\infty \frac{1}{(\#_\lambda j)!},
    \end{split}
  \end{equation}
  where $\#_\lambda j$ is the number of parts $j$ in partition $\lambda$ and $l(\lambda)$ is the length of partition $\lambda$.
  \\
  \\
The $q$-integral is defined as
\begin{equation} \label{qintegral}
\int_{-\nu}^\nu  d_q x  g(x) = (1-q)\sum_{n=0}^\infty  \nu q^n [g(\nu q^n)+ g(-\nu q^n)]
\end{equation}
with $\nu = (1-q)^{-\frac{1}{2}}$. For a $q$-shifted function $x^\dagger g(x)= g(x/q)$, provided $g(x)$ vanish at the boundary the $q$-shift can be rescaled away
\begin{equation} \label{qintegralshift}
\int\limits_{-\nu}^{\nu} g(x/q) \ d_q x = q \int\limits_{-\nu}^{\nu} g(x) \ d_q x .
\end{equation}
One can also produce the $q$-integral by insertion of the operator $\mathcal{O}_\nu(x_i)$,
\begin{equation}
\mathcal{O}_\nu (x_i) = (1-q)\Bigg\{ -\frac{(q;q)_\infty ^2}{\theta (q^\lambda;q)} \left[ f_\lambda(x_i/\nu;q)-f_\lambda(-x_i/\nu;q) \right] \Bigg \} x_i
\end{equation}
with
\begin{equation}
f_\lambda(x;q) = x^\lambda \ \frac{\theta(q^\lambda x;q)}{\theta(x;q)}~,
\end{equation}
which has the desired pole structure provided the choice of contours in figure \ref{contour_nu}. This can be seen using \eqref{qintegral}:
\begin{equation} \label{qintegralpoles}
\begin{split}
\oint_{C_j} \prod_{j=1}^N \frac{dx_j}{x_j} \prod_{i=1}^N \mathcal{O}_\nu (x_i) g(\underline{x}) = &  \oint_{C_j} \prod_{j=1}^N \frac{dx_j}{x_j} \prod_{i=1}^N (1-q)\Bigg\{ -\frac{(q;q)_\infty ^2}{\theta (q^\lambda;q)} \left[ f_\lambda(x_i/\nu;q)-f_\lambda(-x_i/\nu;q) \right] \Bigg \}x_i g(\underline{x}) \\
= & (1-q)^N \sum_{n=0}\nu q^n \left[ g(\nu q^n) + g(-\nu q^n) \right ] \\
= & \int_{-\nu}^\nu \prod_{i=1}^N d_q x_i g(\underline{x}).
\end{split}
\end{equation}
The $q$-exponent has (at least) two definitions
\begin{equation} \label{qexponent}
\begin{split}
e_q(x_i) =  & \sum_{n=0}^\infty \frac{x_i^n}{[n]_q!} \\
= & \prod_{k=0}^\infty (1-(1-q)q^kx_i)^{-1}.
\end{split}
\end{equation}

\section{Verification of starting relation} \label{startingrelation}
\subsection{First starting relation}
The starting relation \eqref{eq:startingrelation}
\begin{equation}
  \sum\limits_{i = 1}^{N} \ x_i^{\dagger} G_i(\underline{x}) F(\underline{x}) = \dfrac{F(\underline{x})}{(t-1)}
  \sum_{i=1}^N\Res_{\omega=q/x_i} \dfrac{d \omega}{\omega - z} \frac{c_q(1/\omega)}{c_q(q/\omega)} \prod\limits_{j = 1}^{N} \dfrac{q - t \omega x_j}{q - \omega x_j}  \prod\limits_{k = 1}^{M} \dfrac{\omega - t y_k}{\omega - y_k}
\end{equation}
can be checked as follows. For the left hand side
\begin{equation}
\begin{split}
\sum_{i=1}^N x_i^\dagger G_i(\underline{x}) F(\underline{x}) = & \sum_{i=1}^N \frac{1}{1-zx_i/q} \prod_{j \neq i} \frac{x_j-tx_i/q}{x_j-x_i/q} \prod_{i \neq l} \frac{(x_i/(qx_l);q)_\infty }{(tx_i/(qx_l);q)_\infty}\prod_{k \neq l, k \neq i } \frac{(x_k/x_l;q)_\infty }{(tx_k/x_l;q)_\infty}\\
& \times \prod_{k \neq i} \frac{(qx_k/x_i;q)_\infty }{(qtx_k/x_i;q)_\infty}\prod_{k \neq l,l\neq i } \frac{(x_k/x_l;q)_\infty }{(tx_k/x_l;q)_\infty} e^{\sum_{k=1}^\infty t_k \left( (x_i/q)^k+ \sum_{j=1, j\neq i}^N x_j^k \right)} \\
& \times c_q(x_i/q)\prod_{j=1, j \neq i}^N c_q(x_j)\\
= & F(\underline{x})\sum_{i=1}^N \frac{1}{1-zx_i/q} \prod_{j \neq i} \frac{(1-tx_j/x_i)}{(1-x_j/x_i)}e^{\sum_{k=1}^N t_k x_i^k \frac{(1-q^k)}{q^k}} \frac{c_q(x_i/q)}{c_q(x_i)}~.
\end{split}
\end{equation}
Whereas the right hand side becomes
\begin{equation}
\begin{split}
\frac{F(\underline{x})}{t-1} & \sum_{i=1}^N \Res_{\omega=q/x_i} \frac{d \omega}{\omega - z} \frac{c_q(1/\omega)}{c_q(q/\omega)}\prod_{j = 1}^{N} \dfrac{q - t \omega x_j}{q - \omega x_j} \prod\limits_{k = 1}^{M} \frac{\omega - t y_k}{\omega - y_k} \\
= & F(\underline{x}) \sum _{i=1}^N  \Bigg \{ \frac{1}{(1-zx_i/q)} \frac{c_q(x_i/q)}{c_q(x_i)} \frac{\prod_{k \neq i}^N (1-tx_k/x_i)}{\prod_{l \neq i}^N (1-x_l/x_i)} e^{\sum_{k=1}^N t_k x_i^k \frac{(1-q^k)}{q^k}}\Bigg \}
\end{split}
\end{equation}
thus agreeing with the left hand side.

\subsection{Second starting relation}
The second starting relation
\begin{equation}
  \sum\limits_{i = 1}^{N} \ x_i^{\dagger} \tilde{G}_i(\underline{x}) F(\underline{x}) = \dfrac{F(\underline{x})}{(t-1)}
  \sum_{i=1}^N \Res_{\omega=q/x_i} d \omega \frac{c_q(1/\omega)}{c_q(q/\omega)} \prod\limits_{j = 1}^{N} \dfrac{q - t \omega x_j}{q - \omega x_j}  \prod\limits_{k = 1}^{M} \dfrac{\omega - t y_k}{\omega - y_k}
\end{equation}
with
\begin{equation}
\tilde{G}_i(\underline{x}) = \dfrac{1}{x_i} \ \prod\limits_{j \neq i} \dfrac{x_j - t x_i}{x_j - x_i}
\end{equation}
can be verified since the left hand side
\begin{equation}
\begin{split}
\sum_{i=1}^N x_i^\dagger \tilde{G}_i(\underline{x}) F(\underline{x}) = & \sum_{i=1}^N \frac{1}{x_i/q} \prod_{j \neq i} \frac{x_j-tx_i/q}{x_j-x_i/q} \prod_{i \neq l} \frac{(x_i/(qx_l);q)_\infty }{(tx_i/(qx_l);q)_\infty}\prod_{k \neq l, k \neq i } \frac{(x_k/x_l;q)_\infty }{(tx_k/x_l;q)_\infty}\\
& \times \prod_{k \neq i} \frac{(qx_k/x_i;q)_\infty }{(qtx_k/x_i;q)_\infty}\prod_{k \neq l,l\neq i } \frac{(x_k/x_l;q)_\infty }{(tx_k/x_l;q)_\infty} e^{\sum_{k=1}^\infty t_k \left( (x_i/q)^k+ \sum_{j=1, j\neq i}^N x_j^k \right)} \\
& \times c_q(x_i/q)\prod_{j=1, j \neq i}^N c_q(x_j)\\
= & F(\underline{x})\sum_{i=1}^N \frac{1}{x_i/q} \prod_{j \neq i} \frac{(1-tx_j/x_i)}{(1-x_j/x_i)}e^{\sum_{k=1}^N t_k x_i^k \frac{(1-q^k)}{q^k}} \frac{c_q(x_i/q)}{c_q(x_i)}
\end{split}
\end{equation}
matches the right hand side
\begin{equation}
\begin{split}
\dfrac{F(\underline{x})}{(t-1)} & \sum_{i=1}^N \Res_{\omega=q/x_i} d \omega \frac{c_q(1/\omega)}{c_q(q/\omega)} \prod\limits_{j = 1}^{N} \dfrac{q - t \omega x_j}{q - \omega x_j}  \prod\limits_{k = 1}^{M} \dfrac{\omega - t y_k}{\omega - y_k} \\
= &F(\underline{x}) \sum _{i=1}^N  \Bigg \{ \frac{1}{(x_i/q)} \frac{c_q(x_i/q)}{c_q(x_i)} \frac{\prod_{k \neq i}^N (1-tx_k/x_i)}{\prod_{l \neq i}^N (1-x_l/x_i)} e^{\sum_{k=1}^N t_k x_i^k \frac{(1-q^k)}{q^k}}\Bigg \}~.
\end{split}
\end{equation}

\section{Verification of algebraic identity} \label{algebraicidentity}

For the algebraic identity
\begin{equation}
\sum\limits_{i = 1}^{N} \ \dfrac{1}{1 - z x_i} \ \prod\limits_{j \neq i} \dfrac{x_j - t x_i}{x_j - x_i} = \dfrac{1}{1-t} - \dfrac{1}{1-t} \prod\limits_{i = 1}^{N} \dfrac{t - z x_i}{1 - z x_i}
\end{equation}
the proof of this is as follows. Starting by considering the contour integral of the right hand side
\begin{equation}
\oint \Big ( \dfrac{1}{1-t} - \dfrac{1}{1-t} \prod\limits_{i = 1}^{N} \dfrac{t - z x_i}{1 - z x_i} \Big ) dz
\end{equation}
where the contour is a big circle with radius tending to infinity, we see that this has apparent singularities at $z = \infty$ and $z = \frac{1}{x_i}$. But taking the limit $z \rightarrow \infty$ (or equivalently $\alpha = \frac{1}{z} \rightarrow 0$)  we see that
\begin{equation}
\begin{split}
\lim_{z \rightarrow \infty} \Big ( \dfrac{1}{1-t} - \dfrac{1}{1-t} \prod\limits_{i = 1}^{N} \dfrac{t - z x_i}{1 - z x_i} \Big ) = & \lim_{\alpha \rightarrow 0} \Big ( \dfrac{1}{1-t} - \dfrac{1}{1-t} \prod\limits_{i = 1}^{N} \dfrac{t \alpha -  x_i}{\alpha - x_i} \Big ) \\
= &  \dfrac{1}{1-t} - \dfrac{1}{1-t} \\
= & 0
\end{split}
\end{equation}
and thus there is no contribution from the singularity at $z = \infty$. Then looking at the other residue at $z = \frac{1}{x_i}$ we see that
\begin{equation}
\begin{split}
\textrm{Res}(z=\frac{1}{x_i})=& \frac{\prod_{j=1}^N (t-\frac{1}{x_i}x_j)}{\prod_{j \neq i}(1-\frac{1}{x_i}x_j)} \\
= & (t-1)\prod_{j \neq i}\frac{ (x_j-tx_i)}{(x_j-x_i)}
\end{split}
\end{equation}
which matches the product on the left hand side. To reproduce this singularity structure  of the right hand side, we need to take this term and multiply by $\sum\limits_{i = 1}^{N} \ \dfrac{1}{1 - z x_i}$, thus generating the algebraic identity.

\section{Contour change} \label{changecontour}
\subsection{First constraint}
\begin{figure}[h]
\includegraphics[scale=0.25]{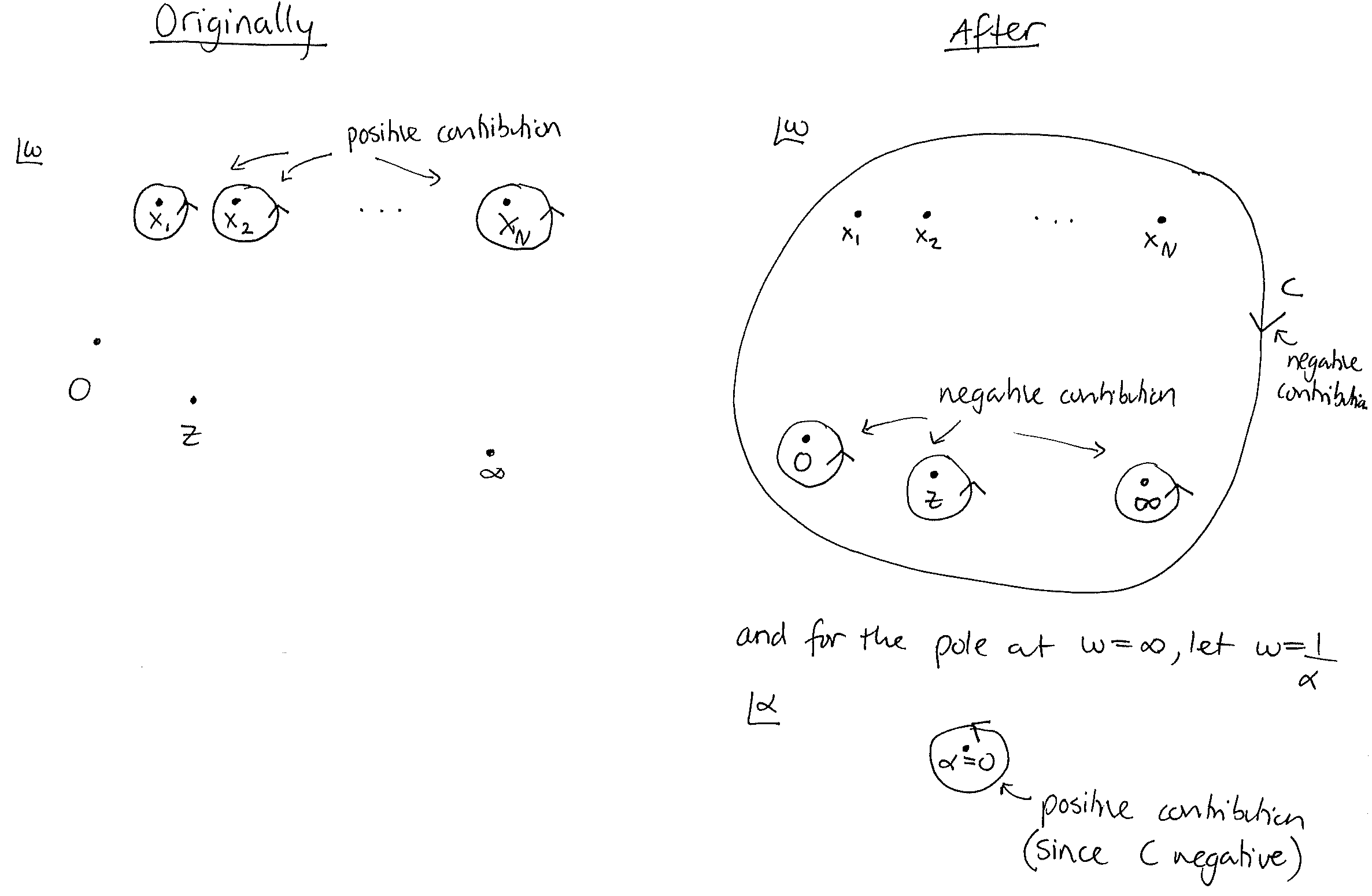}
\centering
\caption{Schematic illustration of the poles appearing in \eqref{eqpoles}} \label{poles}
\end{figure}

In
\begin{equation} \label{eqpoles}
\begin{split}
\oint \prod_{j=1}^N \frac{dx_j}{x_j} & \bigg \{ \frac{1}{1-t} - \frac{1}{1-t} \prod_{i = 1}^{N} \frac{t - z x_i}{1 - z x_i}\bigg \}F(\underline{x}) \\
= & \frac{1}{(t-1)} \oint \prod_{j=1}^N \frac{dx_j}{x_j}   F(\underline{x})\oint \dfrac{d \omega}{\omega - z} \frac{c_q(1/\omega)}{c_q(q/\omega)} \prod\limits_{k = 1}^{N} \dfrac{q - t \omega x_k}{q - \omega x_k}  \prod\limits_{i = 1}^{M} \dfrac{\omega - t y_i}{\omega - y_i}~,
\end{split}
\end{equation}
we then change the order of the $x_i$-integration and the $\omega$-integration on the right hand side so that the poles at $\omega = q/x_i$ are not relevant. Instead the residues at $\omega = z, y_i, 0$ and $\infty$ contribute with opposite sign. (The contour integral over all poles can be moved since we are on a sphere and thus the result is zero. Therefore if we trade the residues at $\omega = q/x_i$ the residues at $\omega = z, y_i, 0$ and $\infty$ will contribute with a minus sign.) This is illustrated in figure \ref{poles}.
\\
\\
Therefore
\begin{equation}
\begin{split}
 \frac{1}{(t-1)}  & \oint \frac{d \omega}{\omega - z} \frac{c_q(1/\omega)}{c_q(q/\omega)} \prod_{i = 1}^{M} \frac{\omega - t y_i}{\omega - y_i} \oint \prod_{j=1}^N \frac{dx_j}{x_j} F(\underline{x}) \prod_{k = 1}^{N} \frac{q - t \omega x_k}{q - \omega x_k} \\
& = - \textrm{Res}(\omega=z)- \sum_{i=1}^M \textrm{Res}(\omega=y_i)- \textrm{Res}(\omega=0)- \textrm{Res}(\omega=\infty)~.
\end{split}
\end{equation}
Now the residues at $\omega = y_i$ can be expanded in formal series since
\begin{equation}
\begin{split}
\prod_{i=1}^N \frac{1}{\omega -y_i} = \frac{1}{\omega^N} \exp \Big (  \sum_{k=1}^\infty \frac{(1-q^k)t_k}{(1-t^k)\omega^k} \Big )
\end{split}
\end{equation}
where in the last line we have used the Miwa variable $t_k = \frac{(1-t^k)}{k(1-q^k)}\sum_i y_i^k$. So these singularities do not contribute to the residue because we are interested in coefficients of $\{t_k\}$.
\\
\\
Similarly we expect the relation \eqref{eqpoles} to hold at each order in a expansion around $z\rightarrow 0$. Thus there is no residue contribution at $\omega = z$ either. Then for the residue at $\omega=\infty$ we consider
\begin{equation}
\begin{split}
\frac{1}{(t-1)} \oint_{\omega=\infty} \Bigg \{ \frac{d \omega}{\omega - z}  \frac{c_q(1/\omega)}{c_q(q/\omega)} \prod_{i = 1}^{M} \frac{\omega - t y_i}{\omega - y_i}  \oint \prod_{j=1}^N  \frac{dx_j}{x_j} F(\underline{x}) \prod\limits_{k = 1}^{N} \frac{q - t \omega x_k}{q - \omega x_k}  \Bigg \}
\end{split}
\end{equation}
and then use the parametrization $\omega = \frac{1}{\alpha}$ with $d \omega = -\frac{1}{\alpha^2}d\alpha$ and instead take the limit $\alpha \rightarrow 0$. Now the contour will be anticlockwise but from the picture above we see that
$$
-\textrm{Res}(\omega=\infty) = -\left(+ \textrm{Res}(\alpha =0) \right)
$$
thus contributing with the same sign. Then we have
\begin{equation}\label{contouralpha}
\begin{split}
\frac{1}{(t-1)} &  \oint_{\alpha = 0} \Bigg \{ \frac{d \alpha}{(-\alpha^2)(\frac{1}{\alpha} - z)}  \frac{c_q(\alpha)}{c_q(q\alpha )} \prod_{i = 1}^{M} \frac{\frac{1}{\alpha} - t y_i}{\frac{1}{\alpha} - y_i} \oint \prod_{j=1}^N \frac{dx_j}{x_j}  F(\underline{x}) \prod\limits_{k = 1}^{N} \frac{q -  \frac{tx_k}{\alpha} }{q - \frac{x_k}{\alpha} }   \Bigg \} \\
= &- \frac{1}{(t-1)}  \oint_{\alpha = 0} \Bigg \{ \frac{d \alpha}{\alpha(1 - z\alpha)}  \frac{c_q(\alpha)}{c_q(q\alpha )}  \prod_{i = 1}^{M} \frac{1 - t y_i\alpha}{1 - y_i\alpha} \oint  \prod_{j=1}^N \frac{dx_j}{x_j}  F(\underline{x}) \prod\limits_{k = 1}^{N} \frac{q \alpha-  tx_k}{q\alpha - x_k }   \Bigg \} \\
= & - \frac{t^N}{(t-1)} \left. \frac{c_q(\alpha)}{c_q(q\alpha )} \right|_{\alpha =0} \oint \prod_{i=1}^N \frac{dx_i}{x_i}  F(\underline{x})~.
\end{split}
\end{equation}
Here we assumed $\left. \frac{c_q(\alpha)}{c_q(q\alpha )} \right|_{\alpha =0} \sim \alpha^0$.  This assumption is reasonable because if the ratio produced positive powers of $\alpha$ then the integrals would be vanishing, and if there were negative powers of $\alpha$ then the equation \eqref{contouralpha} would get powers in $z$ which would destroy the $q$-Virasoro constraints. Thus we can impose that for the $q$-Virasoro constraints to hold we must have $\left. \frac{c_q(\alpha)}{c_q(q\alpha )} \right|_{\alpha =0} \sim \alpha^0.$
\\
\\
Then the equation becomes
\begin{equation}
\begin{split}
\frac{1}{(t-1)} &\oint \prod_{i=1}^N \frac{dx_i}{x_i}   F(\underline{x})\oint \dfrac{d \omega}{\omega - z} \frac{c_q(1/\omega)}{c_q(q/\omega)} \prod\limits_{j = 1}^{N} \dfrac{q - t \omega x_j}{q - \omega x_j}  \prod\limits_{k = 1}^{M} \dfrac{\omega - t y_k}{\omega - y_k}\\
= & -\left( - \frac{t^N}{(t-1)} \left. \frac{c_q(\alpha)}{c_q(q\alpha )} \right|_{\alpha =0} \oint \prod_{i=1}^N \frac{dx_i}{x_i}  F(\underline{x})\right)\\
&-\frac{1}{(t-1)} \oint_{\omega=0} \dfrac{d \omega}{\omega - z} \frac{c_q(1/\omega)}{c_q(q/\omega)} \prod\limits_{i = 1}^{M} \dfrac{\omega - t y_i}{\omega - y_i} \oint \prod_{j=1}^N \frac{dx_j}{x_j} F(\underline{x}) \prod\limits_{k = 1}^{N} \dfrac{q - t \omega x_k}{q - \omega x_k}~.
\end{split}
\end{equation}
\subsection{Second constraint}
We have
\begin{equation}
\begin{split}
\oint \prod_{j=1}^N \frac{dx_j}{x_j} & \bigg \{ \sum\limits_{i = 1}^{N} \dfrac{1}{ x_i} \bigg \}F(\underline{x}) \\
= & \frac{1}{(t-1)} \oint \prod_{i=1}^N \frac{dx_i}{x_i}   F(\underline{x})\oint d \omega \frac{c_q(1/\omega)}{c_q(q/\omega)} \prod\limits_{j = 1}^{N} \dfrac{q - t \omega x_j}{q - \omega x_j}  \prod\limits_{k = 1}^{M} \dfrac{\omega - t y_k}{\omega - y_k}~.
\end{split}
\end{equation}
Changing the order of integration similar to the first constraint equation , and considering the residue at $\omega = \infty$ of the right hand side using the parametrization $\omega = \frac{1}{\alpha}$ with $d \omega = -\frac{1}{\alpha^2} d\alpha$ and instead taking the limit $\alpha \rightarrow 0$,
\begin{equation}
\begin{split}
\frac{1}{(t-1)} & \oint_{\alpha=0} \frac{d \alpha}{-\alpha^2} \frac{c_q(\alpha)}{c_q(q\alpha)} \oint \prod_{i=1}^N \frac{dx_i}{x_i}  F(\underline{x}) \prod\limits_{j = 1}^{N} \dfrac{q - \frac{t x_j}{\alpha} }{q - \frac{x_j}{\alpha} }  \prod\limits_{k = 1}^{M} \dfrac{\frac{1}{\alpha} - t y_k}{\frac{1}{\alpha} - y_k} \\
= & \frac{1}{(t-1)} \oint_{\alpha=0} \frac{d \alpha}{-\alpha^2} \frac{c_q(\alpha)}{c_q(q\alpha)} \oint \prod_{i=1}^N \frac{dx_i}{x_i}   F(\underline{x}) \prod\limits_{j = 1}^{N} \dfrac{q - \frac{t x_j}{\alpha} }{q - \frac{x_j}{\alpha} }  \prod\limits_{k = 1}^{M} \dfrac{\frac{1}{\alpha} - t y_k}{\frac{1}{\alpha} - y_k} \\
= & -\frac{1}{(t-1)} \left. \frac{c_q(\alpha)}{c_q(q\alpha)} \right|_{\alpha=0} \oint \prod_{i=1}^N \frac{dx_i}{x_i}   F(\underline{x}) t^N \Bigg\{ (1-t)\sum_{j=1}^M y_j +q(1-\frac{1}{t})\sum_{j=1}^N\frac{1}{x_j}\Bigg \}
\end{split}
\end{equation}
again assuming $\left. \frac{c_q(\alpha)}{c_q(q\alpha )} \right|_{\alpha =0} \sim \alpha^0$. Using the above the equation becomes
\begin{equation}
\begin{split}
\oint \prod_{j=1}^N \frac{dx_j}{x_j} & \bigg \{ \sum\limits_{i = 1}^{N} \dfrac{1}{ x_i} \bigg \}F(\underline{x}) \\
= & -\Bigg( -\frac{1}{(t-1)} \left. \frac{c_q(\alpha)}{c_q(q\alpha)} \right|_{\alpha=0} \oint \prod_{i=1}^N \frac{dx_i}{x_i}   F(\underline{x}) t^N \Bigg\{ (1-t)\sum_{j=1}^M y_j +q(1-\frac{1}{t})\sum_{j=1}^N\frac{1}{x_j}\Bigg \} \Bigg) \\
&-\frac{1}{(t-1)} \oint \prod_{i=1}^N \frac{dx_i}{x_i}   F(\underline{x})\oint_{\omega=0} d \omega \frac{c_q(1/\omega)}{c_q(q/\omega)} \prod\limits_{j = 1}^{N} \dfrac{q - t \omega x_j}{q - \omega x_j}  \prod\limits_{k = 1}^{M} \dfrac{\omega - t y_k}{\omega - y_k}~.
\end{split}
\end{equation}
\section{Action of $T_0$ and $T_{-1}$ from free field representation} \label{FreeField}
The eigenvalue equations for $T_0$ and $T_{-1}$ can also be seen from the free field representation. Starting with $T_{0}$ we use the explicit form of the generators $T_n$ of \cite{Nedelin:2016gwu} (reproduced in \eqref{generators}) using which we have
\begin{equation}
T_{0}=\sum_{\sigma = \pm 1} q^{\sigma \frac{\sqrt{\beta}}{2}\mathsf{P}}p^{\frac{\sigma}{2}}\sum_{k \geq 0}s_{k}(\{A_{-k}^{(\sigma)}\})s_{k}(\{A_{k}^{(\sigma)}\})~.
\end{equation}
Then
\begin{equation}
\begin{split}
T_0 \mathcal{Z}\{t_k \} = & \sum_{\sigma = \pm 1} q^{\sigma \frac{\sqrt{\beta}}{2}\mathsf{P}}p^{\frac{\sigma}{2}}\sum_{k \geq 0}s_{k}(\{A_{-k}^{(\sigma)}\})s_{k}(\{A_{k}^{(\sigma)}\}) \mathcal{Z}\{t_k \} \\
= & \sum_{\sigma = \pm 1} p^{\frac{\sigma}{2}}q^{\sigma \frac{\sqrt{\beta}\alpha}{2}}s_{0}(\{A_{0}^{(\sigma)}\})s_{0}(\{A_{0}^{(\sigma)}\}) \mathcal{Z}\{t_k \} \\
= & \left( p^{\frac{1}{2}}q^{\frac{\sqrt{\beta}\alpha}{2}}+p^{-\frac{1}{2}}q^{- \frac{\sqrt{\beta}\alpha}{2}} \right) \mathcal{Z}\{t_k \} \\
= & \left( p^{\frac{1}{2}}+p^{-\frac{1}{2}} \right) \mathcal{Z}\{t_k \} ~,
\end{split}
\end{equation}
where in the second line the sum over $k$ drops out because all the positive modes in $A_{k>0}$ annihilate the vacuum, and in the last line we choose the subset of zero momentum states with $\alpha =0$. 
\\
\\
For $T_{-1}$, we again use the explicit form of the generators using which we have
\begin{equation}
T_{-1}=\sum_{\sigma = \pm 1} q^{\sigma \frac{\sqrt{\beta}}{2}\mathsf{P}}p^{\frac{\sigma}{2}}\sum_{k \geq 0}s_{k+1}(\{A_{-1-k}^{(\sigma)}\})s_{k}(\{A_{k}^{(\sigma)}\})~.
\end{equation}
Then
\begin{equation}
\begin{split}
T_{-1} \mathcal{Z}\{t_k \}=& \sum_{\sigma = \pm 1} q^{\sigma \frac{\sqrt{\beta}}{2}\mathsf{P}}p^{\frac{\sigma}{2}}\sum_{k \geq 0}s_{k+1}(\{A_{-1-k}^{(\sigma)}\})s_{k}(\{A_{k}^{(\sigma)}\})\mathcal{Z}\{t_k \} \\
= & \sum_{\sigma = \pm 1} q^{\sigma \frac{\sqrt{\beta}\alpha}{2}}p^{\frac{\sigma}{2}}s_{1}(\{A_{-1}^{(\sigma)}\})s_{0}(\{A_{0}^{(\sigma)}\})\mathcal{Z}\{t_k \} \\
= & \sum_{\sigma = \pm 1} q^{\sigma \frac{\sqrt{\beta}\alpha}{2}}p^{\frac{\sigma}{2}}s_{1}(\{A_{-1}^{(\sigma)}\})\mathcal{Z}\{t_k \} \\
= &\left\{  p^{\frac{1}{2}}\frac{\mathsf{a}_{-1}}{(1+p)}q^{\frac{\sqrt{\beta}\alpha}{2}}+p^{-\frac{1}{2}}\left(-\frac{\mathsf{a}_{-1}}{(1+p^{-1})}q^{-\frac{\sqrt{\beta}\alpha}{2}}\right)\right\}\mathcal{Z}\{t_k \}\\
= &\left\{  p^{\frac{1}{2}}\frac{\mathsf{a}_{-1}}{(1+p)}+p^{-\frac{1}{2}}\left(-\frac{\mathsf{a}_{-1}}{(1+p^{-1})}\right)\right\}\mathcal{Z}\{t_k \}\\
= 0 ~,
\end{split}
\end{equation}
where in the last line we set again $\alpha=0$. Thus we recover the two eigenvalue equations
\begin{equation}
\begin{split}
T_0 \mathcal{Z}\{t_k \} = & \left( p^{\frac{1}{2}}+p^{-\frac{1}{2}} \right) \mathcal{Z}\{t_k \} \\
T_{-1} \mathcal{Z}\{t_k \} = & 0
\end{split}
\end{equation}
from the free field representation in the case of zero momentum.

\section{Conflict of interest statement}
On behalf of all authors, the corresponding author states that there is no conflict of interest.

\providecommand{\href}[2]{#2}\begingroup\raggedright

\bibliographystyle{utphys}
\bibliography{Solving_qVirasoro_constraints}{}

\endgroup

\end{document}